\documentclass[10pt,journal,compsoc]{IEEEtran}

\ifCLASSOPTIONcompsoc
  \usepackage[nocompress]{cite}
\else
  \usepackage{cite}
\fi

\ifCLASSINFOpdf
\else
\fi

\usepackage{url}

\usepackage{tikz}
\usepackage{amsmath}
\usepackage{filecontents}
\usepackage{footnote}
\usepackage{listings}
\usepackage{xspace}
\usepackage{booktabs}
\usepackage{algorithm}
\usepackage{algorithmicx}  
\usepackage{algpseudocode}  
\usepackage{multirow}
\usepackage{amssymb}
\usepackage{pifont}
\usepackage[flushleft]{threeparttable}

\usepackage[font=small,labelfont=bf]{caption}

\definecolor{mygreen}{rgb}{0,0.6,0}  
\definecolor{mygray}{rgb}{0.5,0.5,0.5}  
\definecolor{mymauve}{rgb}{0.58,0,0.82} 
\definecolor{torevise}{rgb}{0.6,0.6,0.6}

\newcommand{\pie}[1]{%
    \begin{tikzpicture}
        \draw (0,0) circle (0.75ex);\fill (0.75ex,0) arc (0:(-#1+90):0.75ex) -- (0,0) -- cycle;
        \fill (0.75ex,0) arc (0:(#1-90):0.75ex) -- (0,0) -- cycle;
    \end{tikzpicture}%
}

\def\eg{\emph{e.g.}\xspace}

\def\ie{\emph{i.e.}\xspace}
\def\etal{\emph{et al.}\xspace}

\newcommand{\tool}[1]{{\textsc{NatiDroid}}}

\usepackage{hyperref}

\begin{document}
%
\title{\textsc{NatiDroid}: Cross-Language \\Android Permission Specification}
%
%
%
%

\author{Chaoran~Li,
        Xiao~Chen,
        Ruoxi~Sun,
        Jason~Xue,
        Sheng~Wen,\\
        Muhammad~Ejaz~Ahmed,
        Seyit~Camtepe,
        Yang~Xiang
\IEEEcompsocitemizethanks{\IEEEcompsocthanksitem {C. Li, S. Wen, and Y. Xiang} are with School of Software and Electrical Engineering, Swinburne University of Technology, Hawthorn, VIC 3122, Australia;\protect\\
E-mail: \{chaoranli, swen, yxiang\}@swin.edu.au;

\IEEEcompsocthanksitem X. Chen is with the Department of Software Systems and Cybersecurity, Faculty of IT, Monash University, Clayton, VIC 3800. Australia; E-mail: xiao.chen@monash.edu;

\IEEEcompsocthanksitem {R. Sun, and J. Xue} are with School of Computer Science, University of Adelaide, North Terrace, SA 5005, Australia;\protect\\
E-mail: \{ruoxi.sun, jason.xue\}@adelaide.edu.au;

\IEEEcompsocthanksitem M. E. Ahmed and S. Camtepe are with CSIRO Data61, Australia;\protect\\
E-mail: \{derek.wang, ejaz.ahmed, seyit.camtepe\}@data61.csiro.au;
}

\thanks{Manuscript received April 19, 2005; revised August 26, 2015.}}

%
%

\markboth{Journal of \LaTeX\ Class Files,~Vol.~14, No.~8, August~2015}%
{Shell \MakeLowercase{\textit{et al.}}: Bare Demo of IEEEtran.cls for Computer Society Journals}
%



\IEEEtitleabstractindextext{%
\begin{abstract}
The Android system manages access to sensitive APIs by permission enforcement. An application (app) must declare proper permissions before invoking specific Android APIs. However, there is no official documentation providing the complete list of permission-protected APIs and the corresponding permissions to date. Researchers have spent significant efforts extracting such API protection mapping from the Android API framework, which leverages static code analysis to determine if specific permissions are required before accessing an API. Nevertheless, none of them has attempted to analyze the protection mapping in the native library (\ie, code written in C and C++), an essential component of the Android framework that handles communication with the lower-level hardware, such as cameras and sensors. While the protection mapping can be utilized to detect various security vulnerabilities in Android apps, such as permission over-privilege and component hijacking, imprecise mapping will lead to false results in detecting such security vulnerabilities. To fill this gap, we thereby propose to construct the protection mapping involved in the native libraries of the Android framework to present a complete and accurate specification of Android API protection. We develop a prototype system, named \tool{}, to facilitate the cross-language static analysis to benchmark against two state-of-the-art tools, termed \textsc{Axplorer}~\cite{backes2016demystifying} and \textsc{Arcade}~\cite{aafer2018precise}. We evaluate \tool{} on more than 11,000 Android apps, including system apps from custom Android ROMs and third-party apps from the Google Play. Our \tool{} can identify up to 464 new API-permission mappings, in contrast to the worst-case results derived from both \textsc{Axplorer} and \textsc{Arcade}, where approximately 71\% apps have at least one false positive in permission over-privilege and up to 3.6\%  apps have at least one false negative in component hijacking. Additionally, we identify that 24 components with at least one Native-triggered component hijacking vulnerability are misidentified by two benchmarks. We have disclosed all the potential vulnerabilities detected to the stakeholders. 
 
\end{abstract}

}

\maketitle

\IEEEdisplaynontitleabstractindextext

%
\IEEEpeerreviewmaketitle

\IEEEraisesectionheading{\section{Introduction}\label{sec:introduction}}

\IEEEPARstart{A}{ndroid} protects access to restricted data (\eg, the device identifier) and actions (\eg, making phone calls) through permission enforcement~\cite{au2012pscout}. Such an access control model can protect users against snooping and protect the stability and security of the operating system~\cite{nauman2010apex}. When an Android app attempts to access the restricted resources, a security check is triggered to inspect whether proper permissions are granted. Lack of permission request will prevent access to the resource and further cease the corresponding functionality or even crash the app.
Therefore, it is essential for developers to know the permissions required of the invoked API. Unnecessary required permissions can pose three threats: \textit{i)} Too many required permissions may confuse users. Users suspect that the app has unexpected behaviors, which leads to users uninstalling or unwilling to install the app. \textit{ii)} The permissions required by an app is an important feature in detecting Android malware. Unnecessary permissions will fool the detector and cause a false alarm. \textit{iii)} The app will incur security risks with unnecessary permissions. Once the app contains vulnerabilities that can be injected with tampered code, it is easy for an attacker to thwart user privacy or invoke sensitive APIs. 
Moreover, requesting unnecessary permissions may expand the attack surface and expose the Android operating system to a host of attacks, especially privilege escalation attacks~\cite{felt2011android}. 
Therefore, to safeguard users' privacy and protect the Android ecosystem, Android app developers are suggested to follow the principle of \textit{least privilege}, \ie, requesting a minimum set of permissions required to fulfill the apps' functionality. Unfortunately, Android does not provide official documentation for the permission specifications (\ie, a mapping between APIs and the required permissions), making it difficult for app developers to follow the least privilege rule, and further lead to security vulnerabilities such as component hijacking~\cite{lu2012chex, wu2013impact, jiang2013detecting}.

To address this problem, researchers have been working on developing methods that generate an accurate list, called a protection map, that maps Android APIs to the required permissions. The previous works that provide such protection maps include \textsc{Stowaway}~\cite{felt2011android}, \textsc{PScout}~\cite{au2012pscout}, \textsc{Axplorer}~\cite{backes2016demystifying}, and, most recently, \textsc{Arcade}~\cite{aafer2018precise}. \textsc{Stowaway} empirically determines the permissions required in Android APIs using feedback-directed testing. \textsc{PScout} and \textsc{Axplorer} leverage control-flow reachability analysis on the source code of the Android framework to generate the mapping between APIs and permissions. \textsc{Arcade} proposes a path-sensitive method based on graph abstraction techniques to generate a more precise mapping. Dynamic testing methods (\eg, \textsc{Stowaway}) can accurately map the required permissions to API invocations that they have tested; however, such dynamic approaches suffer from an intrinsic shortcoming of low coverage. The existing static analysis based approaches (\eg, \textsc{PScout}, \textsc{Axplorer}, and \textsc{Arcade}) have better coverage but may lead to imprecise results because of improper modeling of the complicated Android communication mechanisms. 

Specifically, existing works only analyzed the Java API framework in the \emph{Android API Framework}, but overlooked the \emph{C/C++ Native Library} that consists of core Android system components and services (\eg, Camera service, Sensor service). For example, the public method \texttt{openCamera()} in \texttt{CameraManager.java} class implements its permission check (``android.permission.CAMERA'') in the native library \texttt{CameraService.cpp}. Missing native library analysis will mistakenly conclude that the API \texttt{openCamera()} does not require any permissions (one example is detailed in Section~\ref{sec:background}).

While the API-permission protection mapping contributes to identifying security vulnerabilities in Android apps, such as permission \textit{over-privilege}~\cite{felt2011android} (\ie, an app requests additional permissions that are not required) and \textit{component hijacking}~\cite{lu2012chex} (\ie, an app inappropriately exposes a component with sensitive data), the imprecise mapping will lead to false results on detecting such vulnerabilities. Taking the aforementioned case as an example, an app invokes the API \texttt{openCamera()} will need to request the corresponding permission \texttt{android.permission.CAMERA}; however, existing works that do not analyze the native libraries will identify it as a permission over-privilege case, and henceforce, a false positive.

To address the shortcomings of the existing works, we leverage the cross-language analysis on the overall Android system, including both the \emph{Java API Framework} and the \emph{C/C++ Native Libraries}. 
To this end, we analyze the cross-language communication mechanisms on four Android Open Source Projects (AOSP) and summarize two communication models to facilitate the cross-language analysis. We develop a prototype system, \tool{}, and generate Native-triggered (\ie, an Android API whose permission check is implemented in the native library) API-permission mappings in AOSP versions 7.0, 7.1, 8.0, and 8.1, which were chosen to benchmark against prior works~\cite{backes2016demystifying,aafer2018precise} (see detailed discussion in Section~\ref{Sec:discussion-android-version}). In addition to the mappings generated by previous works, \textsc{Axplorer} and \textsc{Arcade} (2,115 and 1,585, respectively, in AOSP 7.0), \tool{} can successfully discover 449 mappings that are not covered previously. 
Note that while most Android APIs are Java methods, fewer of them are C/C++ methods. Nevertheless, these native methods play indispensable roles in the Android system, such as interacting with the hardware layer. We further use the new mappings to detect permission over-privilege and component hijacking vulnerabilities on a large dataset containing more than 11,000 Android apps, including system apps from custom Android ROMs and third-party apps from the Google Play. We identify the worst-case scenario, where approximately 71\% apps with permission over-privilege detected by \textsc{Axplorer} and \textsc{Arcade} are false positives, as well as both \textsc{Axplorer} and \textsc{Arcade} misidentify 3.6\% apps (false negatives) which are vulnerable to hijacking attacks. Additionally, we identify that 24 components with at least one Native-triggered component hijacking vulnerability are misidentified by two benchmarks.

\begin{figure}[t]
  \centering
  \includegraphics[width=\linewidth]{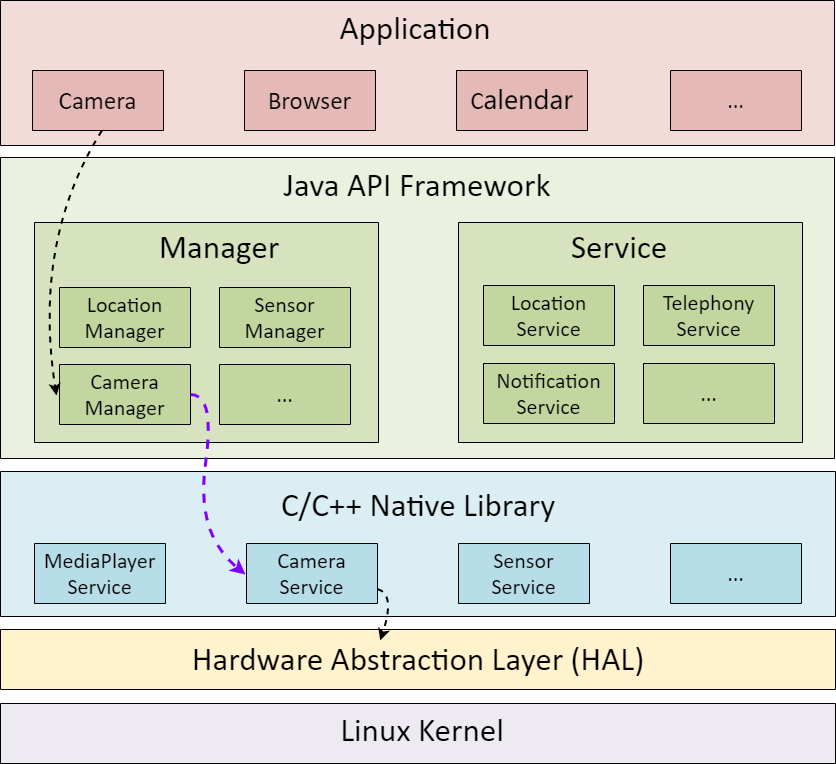}
  \caption{Android software stack}
  \label{structure} 
\end{figure}

In summary, we make the following contributions:
\begin{itemize}
\item We design and implement a prototype system, \tool{}, to facilitate cross-language control-flow analysis on the Android framework. To the best of our knowledge, this is the \textit{first} work to enable cross-language analysis on the Android framework. By incorporating \tool{} with existing Java-side permission mappings (e.g., \textsc{Axplorer} or \textsc{Arcade}), we obtain a complete permission mapping that covers the entire Android system. We make our system and results publicly available to benefit researchers and practitioners.\footnote{The source code is available at \url{https://natidroid.github.io/}\label{opensource:url}.} 

\item We apply \tool{} to extract the permission-API protection mappings from the native libraries on four AOSP versions (7.0, 7.1, 8.0, and 8.1). We show that 12 permissions, including 8 signature and 2 dangerous permissions, are determined to be enforced in native libraries, which are not covered by two state-of-the-art benchmarks, \textsc{Axplorer} and \textsc{Arcade}. 

\item We analyze Android apps for permission over-privilege and component hijacking vulnerabilities at a large scale. Our results show that \tool{} is effective in identifying vulnerable Android apps. We have identified approximately 71\% false positives in terms of the number of the apps with at least one permission over-privileged and up to 3.6\% false negatives in terms of the number of the apps with at least one component hijacking, a worst-case scenario reported by \textsc{Axplorer} and \textsc{Arcade}.
\end{itemize}

We hope that the proposed system, \tool{} in this paper could bridge the gap between Java- and Native-sides analysis (see Figure~\ref{structure}), rendering the static analysis of the overall Android framework to be complete and accurate. 


\section{Background and Motivation}
\label{sec:background}
This section provides background information on how Android OS operates and explains the limitations of the existing static API protection mapping generation techniques that motivate our work.

\noindent\textbf{Android framework.~} 
Android framework consists of the \emph{Java API Framework} layer and \emph{C/C++ Native Library} layer (\ie, the second and the third layers from the top in Figure~\ref{structure}). The \emph{Java API Framework} layer offers Application Programming Interfaces (APIs) written in the Java language for Android app developers to access Android features and functionalities. The Java framework access the device hardware capabilities, such as the camera and sensors, via the \emph{C/C++ Native Library} layer. 
When a Java framework API (\eg, the Camera Manager in Figure~\ref{structure}) invokes a call to access device hardware, the Android system loads corresponding library module (\eg, the Camera Service) for that hardware component. 

\noindent\textbf{Android permission model.~} 
When an app needs to use any of the protected features of an Android device (\eg, accessing the camera, sending an SMS, making a phone call), it must obtain the appropriate permission(s) from the user. When an Android API is called, the Android framework checks whether it is protected by any permission. Most of such permission checks are defined in the \emph{Java API Framework} layer in the Android system, while there are yet a number of them defined in the \emph{C/C++ Native Library} layer.

Existing works~\cite{au2012pscout,backes2016demystifying,aafer2018precise} leverage static analysis on the \emph{Java API Framework} layer of the Android framework to extract the mapping between APIs and corresponding permission checks.
Ignoring the invocation of native libraries miss the permission checks in the native libraries, leading to incomplete mapping results.

\begin{figure}[t]
  \centering
  \includegraphics[width=\linewidth]{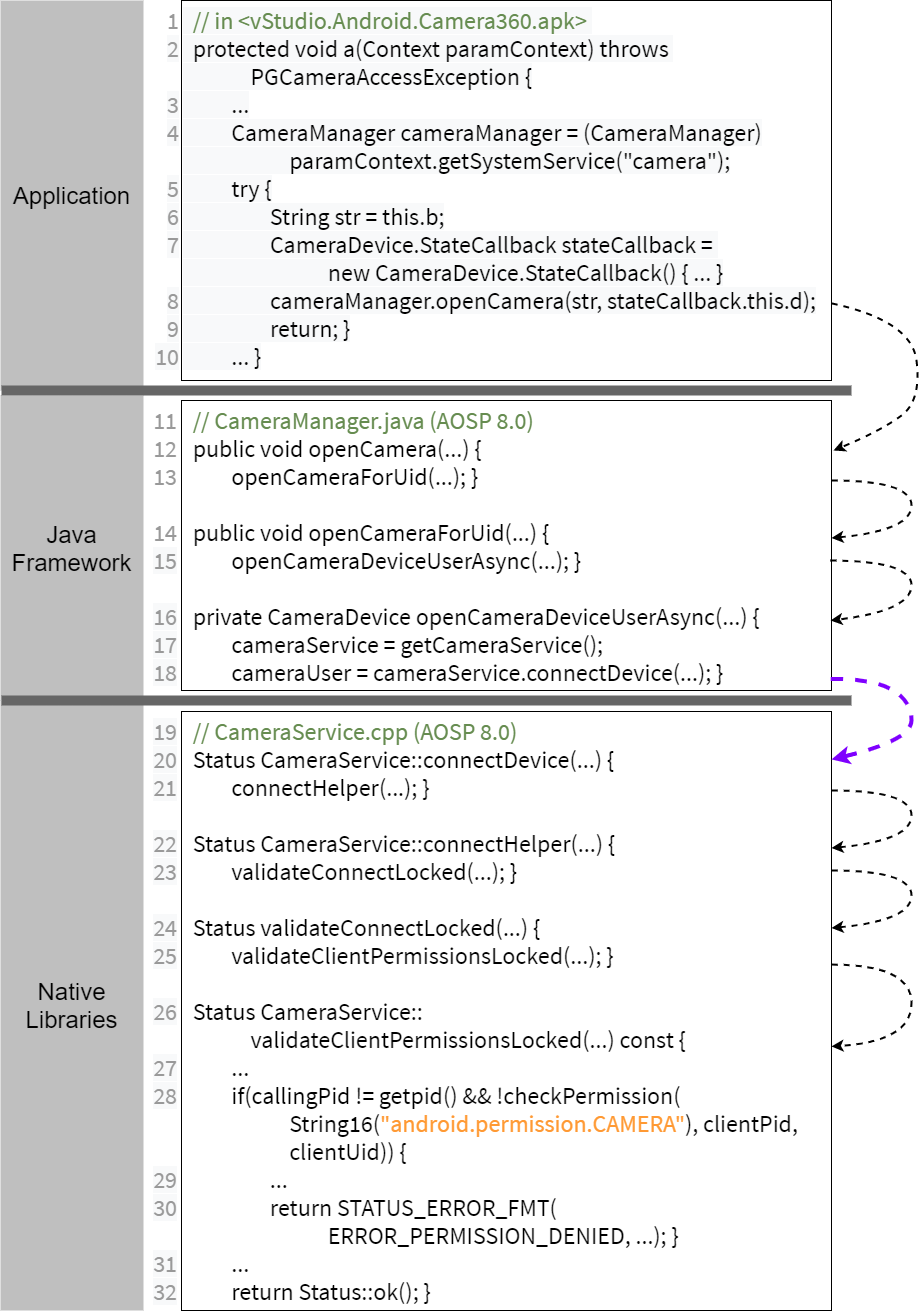}
  \caption{Motivation example derived from a real-world Android app \textit{vStudio.Android.Camera360}. The code is simplified for better illustration.}
  \label{motivation_example} 
\end{figure}

\noindent\textbf{Motivating example.~} 
We further elaborate on our motivation with a real-world example illustrated in Figure~\ref{motivation_example}. The code from lines 1 to 10 is derived from a popular photography app \texttt{vStudio.Android.Camera360} \cite{camera360} on Google Play. The app initialises a \texttt{CameraManager} instance (line 4), and opens the camera instance by invoking \texttt{openCamera()} method (line 8). The invocation chain then traverses along the call path through an Android SDK class \texttt{CameraManager.java} (lines 11 to 18) and a native library \texttt{CameraService.cpp} (lines 19 to 32), and finally triggers a permission check in the native library (line 28). Note that the \texttt{cameraService.connectDevice()} (line 18) communicates with the \texttt{CameraService::connectDevice()} (line 20) in a cross-language way (marked as purple). The security check examines if the method is called by its own process (\ie, \texttt{cameraservice}, hence, no permission is required) or the corresponding permission \texttt{android.permission.CAMERA} is granted. If neither it is called from its own process, nor the \texttt{android.permission.CAMERA} permission is granted, a permission denied error is returned (line 30), which will further prevent \texttt{openCamera()} to be executed. This example implies a protection mapping from the Android API, \texttt{CameraManager.openCamera()}, to its permission protection check, \texttt{\{android.permission.CAMERA || callingPid == getpid()\}}.\footnote{Some \textit{if-then-else} statements are omitted in line 27 of Figure~\ref{motivation_example} for better illustration; consequently, the protection mapping only reflects the simplified code as shown in  Figure~\ref{motivation_example}. The complete protection mapping for \texttt{openCamera()} can be found in our open source repository.}

Unfortunately, as existing works only analyzed the Java source code in the Android framework, they miss the permission checks implemented in the native libraries. For instance, the mapping of the API \texttt{CameraManager.openCamera()} to the permission \texttt{android.permission.CAMERA}, as shown in the example, does not exist in the state-of-the-art works, such as \textsc{PScout}~\cite{au2012pscout}, \textsc{Axplorer}~\cite{backes2016demystifying}, or \textsc{Arcade}~\cite{aafer2018precise}. The incompleteness of the mapping results further introduces false results in detecting security vulnerabilities, such as permission over-privilege~\cite{felt2011android} and component hijacking~\cite{lu2012chex} (detailed in Sections~\ref{subsubsection:overprivilege} and~\ref{subsubsection:hijacking}, respectively).

\begin{figure*}[th!]
  \centering
  \includegraphics[width=\linewidth]{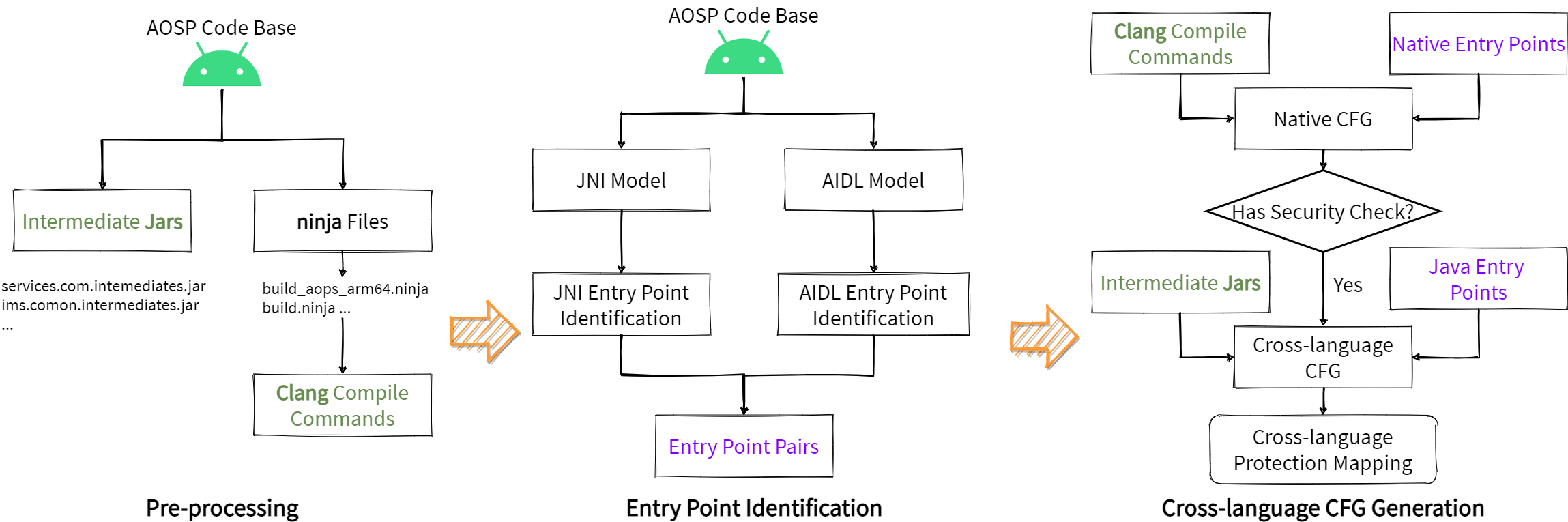}
  \caption{An overview of \textsc{NatiDroid} system}
  \label{high_level} 
\end{figure*}

\section{Approach}
We propose and implement a prototype system, \tool{}, to address the cross-language protection mapping problem that has long been overlooked in previous works. Figure~\ref{high_level} illustrates the overall design of \tool{}. As depicted, \tool{} contains three modules. The \emph{Pre-processing} module prepares the intermediate artifacts for analyzing the Android framework and native libraries, such as intermediate \texttt{.jar} files (for Java-side analysis) and Clang compile commands (for Native-side analysis). The \emph{Entry-points identification} module summarizes two cross-language communication models used in the Android framework, and identifies the entry-points for both Java- and Native-sides analysis. The \emph{Cross-language Control Flow Graph (CFG) analysis} module constructs the cross-language CFG and extracts the permission mapping. 

We propose a complete solution for extracting Native-triggered permission mapping from the Android system. We leverage \textsc{Soot} \cite{soot_} and \textsc{Clang} \cite{clang_} static analysis frameworks, although our solution is also applicable to other static analysis frameworks.
Soot is a popular Java optimization framework for analyzing and visualizing Java and Android apps, which has been widely used in various projects for static analysis~\cite{au2012pscout,fernandes2016security,feng2014apposcopy}. \textsc{Clang} is a lightweight compiler for C language family. We use \textsc{Clang} to transform C/C++ code to the Abstract syntax tree (AST) \cite{ast_}. Additional code for implementing \tool{} consists of approximately 7kLOC.
We detail the design and implementation of each module in the following subsections. 

\subsection{Pre-processing}
\label{Sec:pre-processing}
Due to the complexity and cross-language nature of the Android framework, there is no off-the-shelf tool for static analysis of the Android framework (\ie, \emph{Java API Framework} and \emph{Native Library}) as a whole. \tool{} leverages the \textit{divide-and-conquer} strategy to facilitate the Java- and Native-sides analysis. However, there are still non-trivial tasks to prepare the AOSP codebase for the static analysis. Hence, in this module, we prepare the intermediate artifacts from the AOSP codebase, which are required to enable the Java- and Native-sides analysis. Note that the pre-processing module includes most engineering works, which is not considered our technical contribution. However, it is essential to facilitate the proposed cross-language analysis. 

\noindent\textbf{Java-side analysis preparation.~} 
\tool{}'s Java-side analysis
takes compiled \texttt{.jar} file as input. However,
to maintain the stability of the Android system, some non-SDK class and method bodies are hidden using the \texttt{@hidden} Javadoc tag (\eg, non-SDK Android APIs that may be changed in the future versions without noticing the app developer) during the building of \texttt{android.jar} from source code. The hidden classes and methods only expose the method name, parameters, return values, and minimum set of statements required to handle the invocation, which is not sufficient for constructing a complete CFG. We therefore retain the intermediate output during the compilation, \ie, the intermediate \texttt{.jar} files that have not been combined as \texttt{android.jar}. These intermediate \texttt{.jar} files, such as \texttt{services.com.intermediates.jar}, have the complete class and method information sufficient for facilitating static analysis. 

\noindent\textbf{Native-side analysis preparation.~} 
Before we build the cross-language CFG (cf. Section \ref{sec:module3}), we leverage Clang to transform C/C++ source code to AST.
A complete set of Clang compile command is required to enable the static analysis, however, is not provided in Android documentation. Android uses the ninja to build system~\cite{ninja}. During the compilation process, the \texttt{.ninja} files containing ninja build commands are generated by the compiler. However, the commands obtained from \texttt{.ninja} files consist of file operations and a mixture of GNU Compiler Collection (GCC) and Clang commands, which are not compatible with the off-the-shelf Clang-based analyzer. We then develop a system (to 500 LOC) to extract and pre-process the required commands from these files. The functions of the system include merging separated ninja commands and replacing the \textsc{Clang++} commands with Clang commands (\ie, adding C++ headers in Clang command's parameters). 



\begin{figure}[t]
  \centering
  \includegraphics[width=\linewidth]{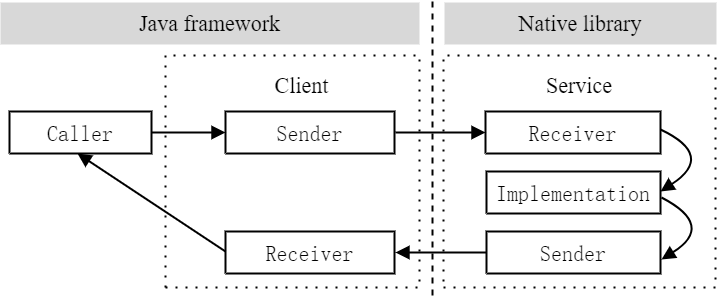}
  \caption{AIDL communication model. The Java-side caller invokes remote method from Native-side. 
  }
  \label{module_1} 
\end{figure}

\subsection{Entry-Points Identification} \label{subsection:entrypoint}
Recall that the overall idea of generating a protection mapping is to examine whether the invocation of an API 
will trigger a permission check in the Android framework (\ie, if there is a permission check node in the CFG starting from the API call). 
Due to the complexity of the Android framework, building a CFG of the overall framework is neither practical nor efficient.   
As \tool{}'s goal is to complement the existing mappings, such as \textsc{Axplorer} and \textsc{Arcade}, by adding the protections whose permission checks are located in the native libraries, we only generate sub-CFGs for the Android APIs that involve cross-language communication. 
The first step to generate sub-CFGs is to identify the entry-points of the graphs (for both Java- and Native-sides). To this end, we first summarize two cross-language communication mechanisms used by Android. 

\begin{figure}[t]
    \centering
    \includegraphics[width=\linewidth]{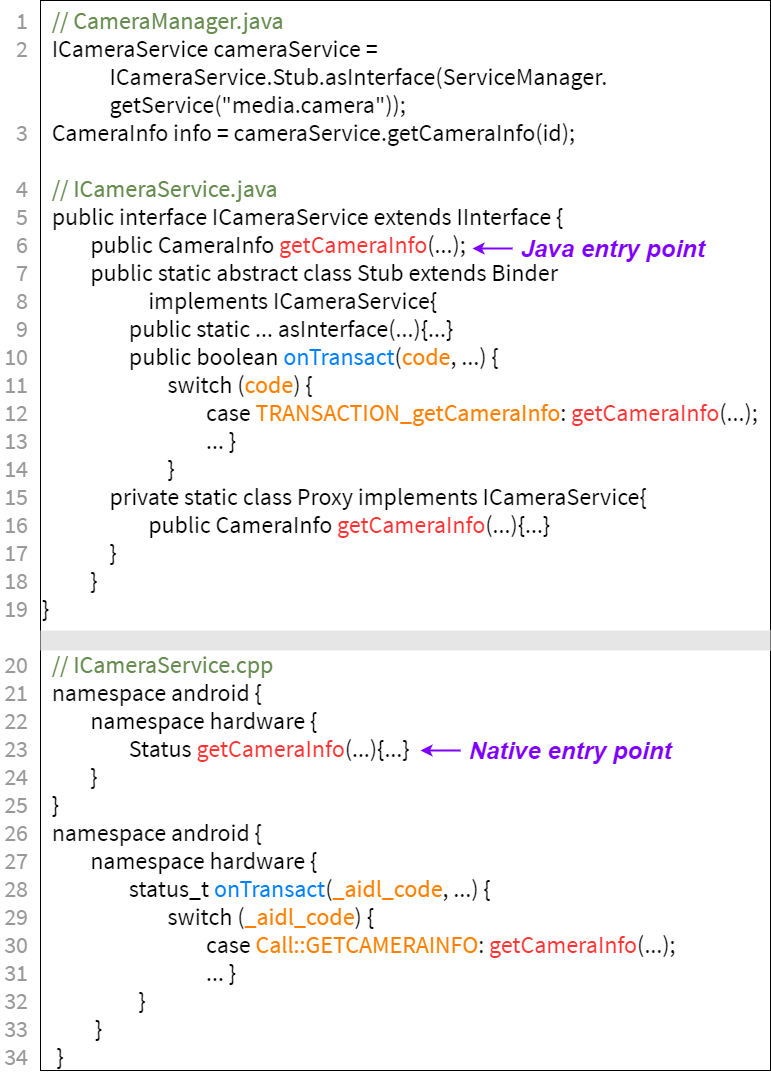}
    \caption{An example of AIDL-based cross-language communication model. The code snippets are simplified for better illustration.}
    \label{code:Client_Server_exmaple}
\end{figure}

\noindent\textbf{AIDL-based communication model.} 
The Android operating system (OS) is based on Linux but does not inherit the Inter-Process Communication (IPC) mechanism of Linux. In Android, resources between processes are not shared. Android Interface Definition Language (AIDL) is one of the IPC mechanisms that Android adopted to enable communication for remote invocation, providing cross-process or even cross-language communication. 
Figure~\ref{module_1} depicts the workflow of AIDL-based client-service model, where the Java framework works as a client requesting service from the native library. AIDL utilizes a pair of Stub/Proxy classes on Java-side to communicate with native libraries. The \texttt{Proxy} is responsible for sending requests to native service and implementing the remote method which invokes the \texttt{transact()} method and communicates with Native-side, while the \texttt{Stub} class, inheriting the \texttt{Binder} class, transforms service \texttt{Binder} and receives the reply from native service using the method \texttt{onTransact()}. The \texttt{transact()} and \texttt{onTransact()} are synchronous, such that a call to \texttt{transact()} does not return until the target has returned from \texttt{onTransact()}. 

On Native-side, it is unnecessary to generate Stub/Proxy pairs, but directly implements the remote method (using the name as same as the remote method on Java-side) to handle the request from Java-side, so that we can always find the receiver \texttt{onTransact()} and the same name remote method as the AIDL sender. 
Through using pairs of \texttt{onTransact()} and \texttt{transact()} methods as sender and receiver on both sides, the communication between Java- and Native-sides is established. 
Therefore, cross-language interaction can be detected via matching the use of AIDL on Java- and Native-sides (\eg, the Stub/Proxy pair, the \texttt{onTransact()} methods, and the implementation of the remote methods and the \texttt{transact()} methods). We can then determine the entry-points for static analysis accordingly. 

An example of AIDL mechanism is shown in Figure~\ref{code:Client_Server_exmaple}, where the method \texttt{getCameraInfo()} uses AIDL to implement the communication between Java and C++. A pair of a sender and a receiver on each side of AIDL (\ie, the client and the service) handles the cross-language communication. The caller (line 3) invokes the method \texttt{getCameraInfo()} which is firstly defined as an interface in line 6 and then implemented in line 16 (the detailed implementation is omitted). In the corresponding native library (lines 21 to 34), the receiver \texttt{onTransact()} handles the request and further invokes the method \texttt{getCameraInfo()} (line 30). The \texttt{getCameraInfo()} method then executes the method implementation and sends the execution result back to the Java-side \texttt{onTransact()} method (line 10), which is further passed back to the caller (line 3). Note that the \texttt{getCameraInfo()} in line 6 and the \texttt{getCameraInfo()} in line 23 are the two interface methods that both invoke the \texttt{transact()} method (omitted) to establish the communication between Java- and Native-sides. The \tool{} will match the pair of such remote methods and recognize them as a pair of Java entry-point and native entry-point. Note that, for the inner-language AIDL communication on both Java- and Native-sides, \tool{} generates CFG with a similar approach, except for the identification of entry-points.

\begin{figure}[t]
  \centering
  \includegraphics[width=\linewidth]{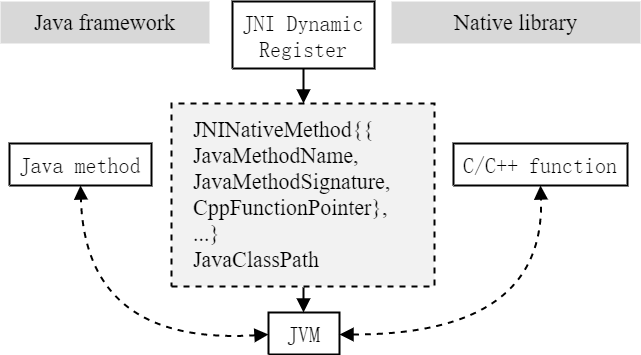}
  \caption{JNI communication model. Java- and Native-sides communicate with JNI.}
  \label{module_2} 
\end{figure}

\noindent\textbf{JNI-based communication model.}
Java Native Interface (JNI) provides a cross-language interface between Java and Native code (written in C/C++). 
This enables developers to write native methods to handle situations when an app cannot be written entirely in the Java programming language, \eg, when the standard Java class library does not support the platform-specific features or program library, such as communication with the underlying hardware. In the JNI framework, native functions are implemented in \texttt{.c} or \texttt{.cpp} files. 
When invoking the function, it passes a \texttt{JNIEnv} pointer, a \texttt{jobject} pointer, and any Java arguments declared by the Java method. 

Figure~\ref{module_2} shows the JNI-based communication model adopted by Android. Android uses the JNI Dynamic Register to link native methods. Different from the AIDL model, the JNI-based communication starts from a registration process. 
When Android starts running, the \texttt{AndroidRuntime} class uses the \texttt{startReg()} method to start the registration of JNI methods, which will further invoke all JNI registration functions implemented in the native libraries. 
The registration functions will register native methods with the corresponding Java class specified by an array of \texttt{JNINativeMethod} structures that contains the Java method name, Java method signature, and pointers to the native function. 
After the registration process, all the \texttt{JNINativeMethod} (on Native-side) is registered and linked to the corresponding Java method in the Java Virtual Machine (JVM). 

\begin{figure}[t]
    \centering
    \includegraphics[width=0.97\linewidth]{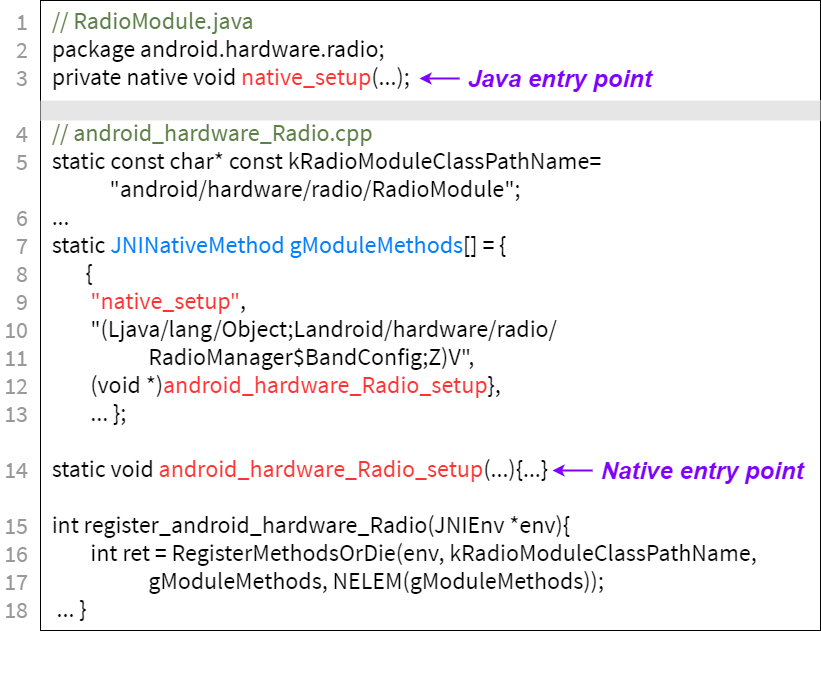}
     \vspace{-3mm}
    \caption{An example of JNI-based cross-language communication. The code snippets are simplified for better illustration.}
    \label{code:JNI_example}
\end{figure}

We further explain the JNI-based communication mechanism with an example given in Figure \ref{code:JNI_example}, which is derived from \texttt{android\_hardware\_Radio.cpp} in AOSP 8.0. 
The method \texttt{register\_android\_hardware\_Radio()} (line 15) is called to register the JNI methods for Radio, with the JNI method information provided in line 16. Specifically, the \texttt{kRadioModuleClassPathName} variable (line 5) declares the containing class name of the Java method, and \texttt{gModuleMethods} (line 7) declares the correspondence between the Java method and the C++ function. The variable \texttt{gModuleMethods} is defined to contain groups of the Java method name (line 9), parameter and returned types (lines 10 to 11), and the pointer of C++ method (line 12). All the information will be dynamically registered in JVM during run-time. 
Finally, the C++ method involved in the cross-language communication is declared in line 14, 
while the involving Java method can be found in the java file \texttt{RadioModule.java} in package \texttt{android.hardware.radio} (lines 2 to 3). 

According to the JNI-based communication mechanism, we extract pairs of entry-points from Java- and Native-sides. Specifically, the JNI methods and the corresponding native methods are vaguely identified by a linear sweep searching of keyword \texttt{RegisterMethodsOrDie}, \texttt{registerNativeMethods} and \texttt{jniRegisterNativeMethods} over the \texttt{.cpp} files at first. Then we extract the class path name (\eg, the \texttt{kRadioModuleClassPathName} in Figure~\ref{code:JNI_example}) and the array of \texttt{JNINativeMethod} structures, from which a pair of entry-points can be located and recognized (\eg, the pair of \texttt{native\_setup()} and \texttt{android\_hardware\_Radio\_setup()}). 

\subsection{Cross-Language Protection Mapping Extraction}
\label{sec:module3}
After the entry-points on both Java- and Native-sides are identified, we further extract the protection mappings from AOSPs. In this section, we first introduce how \tool{} generates CFG from both sides. Based on the CFG, \tool{} then traverses the cross-language Android API call paths and corresponding security checks (\eg, permission checks) to generate the API-permission protection mappings.
\subsubsection{Cross-language CFG Generation}

\renewcommand{\algorithmicensure}{\textbf{Output:}}
\begin{algorithm}[t]
\caption{{\bfseries Constructing Cross-language CFG}}
\textbf{Input:} Android native code and intermediate jar $c$.\\
\textbf{Output:} Cross-language CFG $G$.
\begin{algorithmic}[1]
    \State $G = \emptyset$
    \State $(N_{epj}, N_{epc}) \gets scanEntryPoints(c)$
    \For{$each\ (n_{epj}, n_{epc}) \in (N_{epj}, N_{epc})$}
        \State $(N_c, E_c) \gets NativeCFGGenerator(n_{epc})$
        \If{$hasSecurityCheck(N_c)$}
            \For{$each\ (n_c, e_c) \in (N_c, E_c)$}
                \State $G \gets G \cup (n_c, e_c)$
            \EndFor
            \State $(N_j, E_j) \gets JavaCFGGenerator(n_{epj})$
            \For{$each\ (n_j, e_j) \in (N_j, E_j)$}
                \State $G \gets G \cup (n_j, e_j)$
            \EndFor
            \State $e \gets link(n_{epj}, n_{epc})$
            \State $G \gets G \cup (n_{epj}, e)$
            \State $G \gets G \cup (n_{epc}, e)$
        \EndIf
    \EndFor
\end{algorithmic}
\label{generate_CFG}
\end{algorithm}

Algorithm~\ref{generate_CFG} elaborates the detailed steps involved in generating cross-language CFGs.
After obtaining the entry-point pairs from both Java- and Native-sides (line 2, as detailed in Section~\ref{subsection:entrypoint}), \tool{} first leverages a forward analysis to generate a CFG on the native side from each identified native entry-point (line 4). If the native-side CFG does not contain any security checkpoint (\eg, permission check, \textit{UID} check, and \textit{PID} check), we discard the CFG for computational efficiency (line 5).
Otherwise, \tool{} further utilizes a backward analysis to build a Java-side CFG starting from the paired Java-side entry-point to an Android API. If the reached Android API is further invoked by other Android APIs, we extend the CFG until the API is not called by any other APIs (lines 6 to 12). The CFGs generated from both sides are then connected with the communication models identified in Section~\ref{subsection:entrypoint} (lines 13 to 15).
 
We detail the mechanisms unique to Android that require additional work to handle as follows.

\begin{figure}[t]
    \centering
    \includegraphics[width=0.96\linewidth]{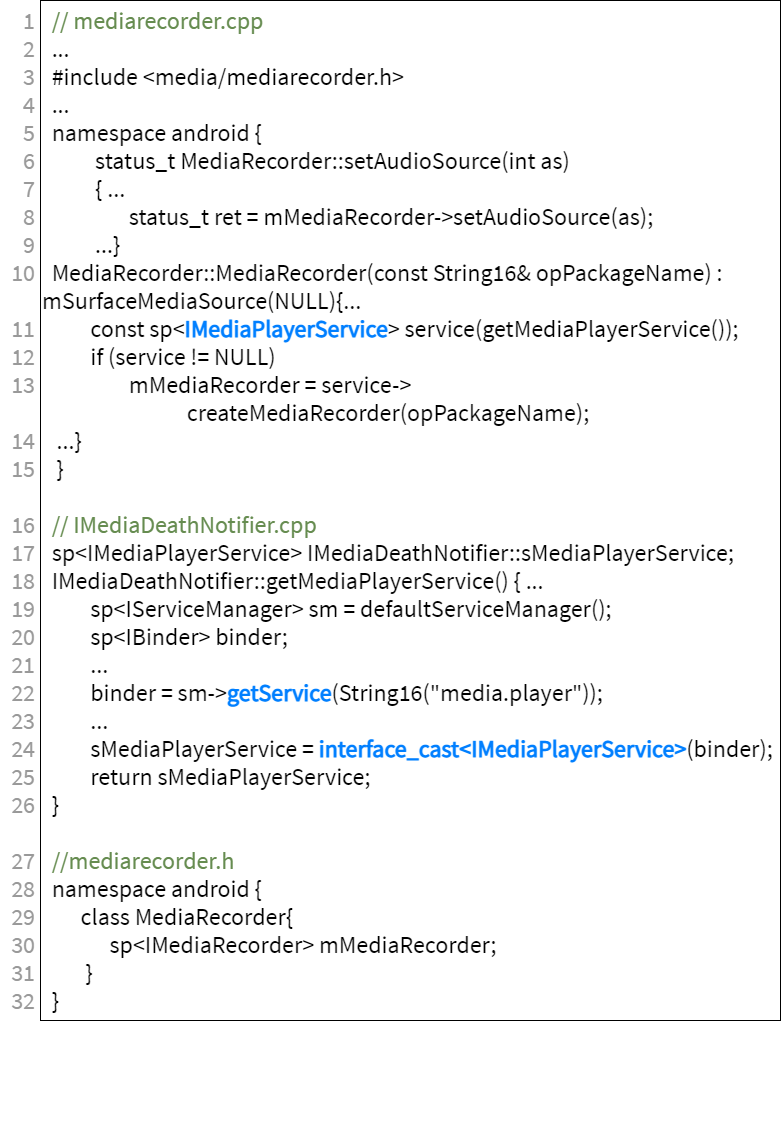}
     \vspace{-10mm}
    \caption{An example of using strong pointer in native library}
    \label{code:native_service}
     \vspace{-3mm}
\end{figure}

\noindent\textbf{Handling the service identifier.~} 
The aforementioned AIDL is often used to invoke remote methods in service. Before the invocation, the service is usually pointed by passing a string to the \texttt{getService} or \texttt{checkService} method, for example, the string \texttt{``media.player''} in line 22 of Figure~\ref{code:native_service}. When building the call graph, we need to handle such remote invocation and identify which class is the identifier string actually pointed to. These services are registered to the system through an \texttt{addService} method (either on Java-side or Native-side). Therefore, we can automatically collect the correspondences between these identifiers and service classes from it. First, we scan the Java and C++ files looking for \texttt{addService} methods. Then the program confirms whether the method is \texttt{ServiceManager.addService} or \texttt{defaultServiceManager->addService}, separately. Once confirmed, the program extracts a pair of service class and the corresponding identifier from the parameters; for example, the string identifier \texttt{``media.player''} in Figure~\ref{code:native_service} will be paired with its service class \texttt{MediaPlayerService}.

\noindent\textbf{Handling Android strong pointer.~} 
Although the strong pointer defines the type variable, the type is not necessary to be restricted. Therefore, we need additional efforts to get what type the strong pointer actually points to from the context. In the case shown in Figure~\ref{code:native_service}, the \texttt{createMediaRecorder()} method of the \texttt{service} object is invoked (line 13). Considering \texttt{createMediaRecorder()} is a member function and the object \texttt{service} is defined as a strong pointer, to determine which \texttt{createMediaRecorder()} method is actually invoked, we need to determine the type of \texttt{service} at first. By tracing the variables, we can find that, in line 25, the \texttt{getMediaPlayerService()} returns variable \texttt{sMediaPlayerService} to \texttt{service}, which is converted from the variable \texttt{binder} in line 24. Further, according to line 22, the \texttt{binder} is the result of \texttt{getService()} which returns a service object in the type of \texttt{MediaPlayerService} (as detailed in the previous paragraph). Therefore, we can determine that the \texttt{createMediaRecorder()} method invoked in line 13 is a method defined in \texttt{MediaPlayerService} class, rather than the \texttt{IMediaPlayerService} class as declared by the strong pointer in line 11.
According to the strong pointer mechanism, when we find that a member function is called, if the object is declared as a strong pointer, the invocation of the member function will be determined automatically. The \tool{} will first trace the statement where an object is assigned to the strong pointer. If it is assigned through mehtod \texttt{getService()}, the type of the object will be determined by the passed service identifier (as detailed in the previous paragraph); otherwise, \tool{} will determine the type of strong pointer according to the variable type returned by the function.

\noindent\textbf{Handling member variables.~}
A class member variable is also possibly assigned by a strong pointer. For example, in line 8 of Figure~\ref{code:native_service}, the \texttt{setAudioSource()} method of the \texttt{mMediaRecorder} object is invoked, but from this line of code we cannot determine the type of variable \texttt{mMediaRecorder} so that we cannot determine which \texttt{setAudioSource()} method has been invoked. By looking up in the referenced header file (from line 27), we find that this variable is a member variable of class \texttt{MediaRecorder}. Therefore, to find out the type of this variable, we can search the entire class looking for the assignment or initialization statement. Note that the assignment should be ignored if the assignment releases the pointer, \eg, pointing the variable to a null pointer. In this case, the variable \texttt{mMediaRecorder} is initialized as the return value of \texttt{createMediaRecorder} in the class constructor (line 13). We have explained how to determine the return type of \texttt{createMediaRecorder} in the previous paragraph, handling Android strong pointers. To implement the process automatically, if a variable cannot be tracked in the local scope, \tool{} will point to the header file to check whether it is a member variable, and thus the tracking scope will be expanded to the entire class.

\subsubsection{Protection Mapping Extraction} 

Algorithm~\ref{extract_maps} shows the pseudo-code of extracting the protection mapping of Android APIs.
After obtaining the cross-language CFG, we resort to a Depth-First Search (DFS) strategy (line 2) to check if there are call paths between Android APIs in the \emph{Java API Framework} layer and security checks in the \emph{Native Library} layer. For each node in the CFG, if it is a Java node, we will collect all its native children (lines 3 to 5) and obtain the security checks defined in its children nodes (lines 7 to 8).
If there are more than one checkpoint on the call trace (\eg, an Android API is protected by multiple permissions), we concatenate them with \textit{AND} operation. As inspired by Aafer \etal~\cite{aafer2018precise}, we also include security checks other than permission enforcement, such as UID and PID checks. If there is more than one Android API along the track, we create a mapping entry for each of them (\ie, all Android APIs along the track have the same security check). Finally, each pair of Java API node and its corresponding security check(s) in Native-side will be added into the protection mapping (line~10).

\begin{algorithm}[t]
\caption{{\bfseries Protection Maps Extraction}}
\textbf{Input:} The cross-language CFG generated in Algorithm 1 $G$.\\
\textbf{Output:} The protection mapping $M$.
\begin{algorithmic}[1]
    \State $M = map(Node,\ Condition)$
    \State $N \gets DFS(G)$
    \For{$each\ n \in N$}
        \If{$isJavaAPINode(n)$}
            \State $K \gets getAllNativeChildren(n)$
            \For{$each\ k \in K$}
                \If{$hasSecurityCheck(k)$}
                    \State $C \gets C \cup getSecurityChecks(k)$
                \EndIf
                \State $M \gets M \cup (n, C)$
            \EndFor
        \EndIf
    \EndFor
\end{algorithmic}
\label{extract_maps}
\end{algorithm}

\begin{table*}[t]
\centering
\caption{Permission-API protection mappings extracted by \tool{}. The results of \textsc{Axplorer} and \textsc{Arcade} are generated from the Android Java Framework, while the results of \tool{} are generated from Android Native Libraries.}
 \label{tab:performance_comparison}  
 \resizebox{\linewidth}{!}{
 \begin{threeparttable}
\begin{tabular}{cccccccccccccccccc}
\toprule
\multicolumn{1}{c}{\multirow{5}{*}{\textbf{Source}}} & \multicolumn{1}{c}{\multirow{5}{*}{\textbf{\begin{tabular}{c}\textbf{Analysis}\\\textbf{Time}\\\textbf{(min)}\end{tabular}}}} & \multicolumn{10}{c}{\textbf{API Framework (Java)}} & \multicolumn{5}{c}{\textbf{Native Library (C/C++)}} \\ 

\cmidrule(r){3-12} \cmidrule(r){13-17}
& & \multicolumn{5}{c}{\textbf{\textsc{Axplorer}}}  & \multicolumn{5}{c}{\textbf{\textsc{Arcade}}} & \multicolumn{5}{c}{\textbf{\textsc{NatiDroid}}}  \\ 

\cmidrule(r){3-7} \cmidrule(r){8-12}  \cmidrule(r){13-17} 
& & \multicolumn{4}{c}{\textbf{\# Permission}} & \multirow{2}{*}{\textbf{\# API Affected}}  & \multicolumn{4}{c}{\textbf{\# Permission}} & \multirow{2}{*}{\textbf{\# API Affected}} & \multicolumn{4}{c}{\textbf{\# Permission}} & \multirow{2}{*}{\textbf{\# API Affected}} \\ 

\cmidrule(r){3-6} \cmidrule(r){8-11} \cmidrule(r){13-16}
& &  
\textbf{\# S} \tnote{1} & \textbf{\# D} & \textbf{\# N} & \textbf{\# Total} & & 
\textbf{\# S} & \textbf{\# D} & \textbf{\# N} & \textbf{\# Total} & & 
\textbf{\# S} & \textbf{\# D} & \textbf{\# N} & \textbf{\# Total} &\\ 

\midrule
\multirow{1}{*}{\textbf{AOSP 7.0}} & 43 & 145 & \textbf{12} & 27 & 184 & 2,115 & 152 & \textbf{11} & 30 & 193 & 1,585 & 5 & \textbf{2} & 2 & 9 & \textbf{449}\\

\multirow{1}{*}{\textbf{AOSP 7.1}} & 45 & 145 & \textbf{12} & 27 & 184 & 2,153 & 152 & \textbf{11} & 30 & 193 & 1,585 & 6 & \textbf{2} & 2 & 10 & \textbf{449}\\
\multirow{1}{*}{\textbf{AOSP 8.0}} & 49 & / & / & / & / & / & / & / & / & / & / & 8 & \textbf{2} & 2 & 12 & \textbf{464}\\
\multirow{1}{*}{\textbf{AOSP 8.1}} & 51 & / & / & / & / & / & / & / & / & / & / & 7 & \textbf{2} & 2 & 11 & \textbf{461}\\
\bottomrule
\end{tabular}
  \begin{tablenotes}
    \item[1] \textbf{S}: Signature permission; \textbf{D}: Dangerous permission; \textbf{N}: Normal permission.
  \end{tablenotes}
\end{threeparttable}
}
\end{table*}

\section{Evaluation}
The main contribution of this work, \tool{}, is to enable the cross-language static analysis of the Android framework. Since there is a number of existing tools~\cite{vallee2010soot, wala} and works~\cite{backes2016demystifying, aafer2018precise} well handling the static analysis of the Java-side of Android framework, \tool{} specifically focuses on the analysis of the cross-language part of the Android framework. We therefore incorporate existing works \textsc{Axplorer}~\cite{backes2016demystifying} and \textsc{Arcade}~\cite{aafer2018precise} for the analysis that does not involve cross-language communication. 

\subsection{Protection Mapping in Android Native Libraries}

We use \tool{} to extract the API protection mapping for four AOSP versions -- 7.0, 7.1, 8.0, and 8.1. We obtained the source code from the official AOSP repository \cite{aosp_codebase}. The experiment was run on a Linux server with Intel (R) Core (TM) i9-9920X CPU @ 3.50GHz and 128 GB RAM.

\begin{table}[t]
\centering
\caption{Permission checks in native libraries}
\label{tab:permission_results}
\resizebox{\linewidth}{!}{
\begin{threeparttable}
\begin{tabular}{@{}lccccl@{}}
\toprule
\multirow{2}{*}{\textbf{Permissions}} & \multicolumn{4}{c}{\textbf{AOSP}} & \multirow{2}{*}{\begin{tabular}{c}\textbf{Protection}\\\textbf{Level}\end{tabular}} \\ 
 & \textbf{7.0} & \textbf{7.1} & \textbf{8.0} & \textbf{8.1} & \\ \midrule
\textbf{ACCESS\_DRM\_CERTIFICATES} & \pie{360} & \pie{360} & \pie{360} & \pie{360} & \textcolor{orange}{Signature} \\
\textbf{ACCESS\_FM\_RADIO} & \pie{90} & \pie{90} & \pie{360} &  \pie{90} & \textcolor{orange}{Signature}\\
\textbf{ACCESS\_SURFACE\_FLINGER} & \pie{360} & \pie{360} & \pie{360} & \pie{360} & \textcolor{orange}{Signature} \\
\textbf{CAPTURE\_AUDIO\_HOTWORD} & \pie{360} & \pie{360} & \pie{360} & \pie{360} & \textcolor{orange}{Signature} \\
\textbf{CONTROL\_WIFI\_DISPLAY} & \pie{90} & \pie{360} & \pie{360} & \pie{360} & \textcolor{orange}{Signature} \\
\textbf{LOCATION\_HARDWARE} & \pie{90} & \pie{90} & \pie{360} & \pie{360} & \textcolor{orange}{Signature} \\
\textbf{MODIFY\_AUDIO\_ROUTING} & \pie{360} & \pie{360} & \pie{360} & \pie{360} & \textcolor{orange}{Signature} \\
\textbf{READ\_FRAME\_BUFFER} & \pie{360} & \pie{360} & \pie{360} & \pie{360} & \textcolor{orange}{Signature} \\
\textbf{RECORD\_AUDIO} & \pie{360} & \pie{360} & \pie{360} & \pie{360} & \textcolor{red}{Dangerous} \\
\textbf{CAMERA} & \pie{360} & \pie{360} & \pie{360} & \pie{360} & \textcolor{red}{Dangerous}\\
\textbf{INTERNET} & \pie{360} & \pie{360} & \pie{360} & \pie{360} & Normal \\
\textbf{MODIFY\_AUDIO\_SETTINGS} & \pie{360} & \pie{360} & \pie{360} & \pie{360} & Normal \\ \bottomrule
\end{tabular}
\begin{tablenotes}
\item \pie{360}: The permission exists in AOSP; \pie{90}: The permission does not exist in AOSP.
\end{tablenotes}
\end{threeparttable}
}
\end{table}

Table~\ref{tab:performance_comparison} presents the statistics of permission-API protection mapping on the four AOSP versions. Since \tool{} only focuses on the Native-triggered permission checks (\ie, on the Native-side), the results of the Java API framework (\ie, on the Java-side) are derived from previous works \textsc{Axplorer} and \textsc{Arcade} \cite{backes2016demystifying,aafer2018precise}.\footnote{Java-side mappings are derived from \url{https://github.com/reddr/axplorer/} and \url{https://arcade-android.github.io/arcade/}. Since the authors of both works only released their mapping results (for AOSP version 7.1 and under) in lieu of the tools, we are unable to obtain the mappings on AOSP 8.0 and 8.1.} As shown in the last two columns of Table~\ref{tab:performance_comparison}, the number of newly identified permissions that are missed in previous works ranges from 9 to 12 in the four AOSP versions. There are 449 to 464 Android APIs associated with these permissions (\ie, invoking these APIs requires the corresponding permissions to be granted), counting up to approximate
30\% of the mappings reported in the previous study, which are overlooked in the previous work that only analyzed the Java-side of the Android framework.

Table~\ref{tab:performance_comparison} reports the breakdown of the mappings based on the permission protection levels. The mapping results contain the permissions in \textit{signature}, \textit{dangerous}, and \textit{normal} levels. \textit{Signature} permissions are only granted if the requesting app is signed with the same certificate as the app that declared the permission. The signature permissions declared in the AOSP will only be granted to apps developed by Android; hence, neither mobile phone vendors (\eg, Samsung, Huawei) nor third-party app developers have access to them. \textit{Normal} permissions refer to the permissions with minimal risk to the system and the users' private data. \textit{Dangerous} permissions are those higher-risk permissions that would give access to private user data or control over the device that can negatively impact the users. Missing these mappings, especially the ones for the dangerous permissions, will lead to false results in detecting security issues of Android apps, such as permission over-privilege and component hijacking problems (detailed in Section~\ref{applications_of_protection_mapping}). Given that \textit{signature} permissions can only be accessed by system apps and \textit{normal} permissions are usually associated with non-sensitive behaviours. 
Thus, the main security and privacy threats to the majority of Android apps (i.e., third-party apps downloaded from official or alternative app stores, which usually have no access to \textit{signature} permissions) are caused by inaccurate mapping of \textit{dangerous} permissions. It is therefore worth highlighting that \tool{} is able to find approximately 17\% dangerous level permission that previous works have missed. \tool{} has identified the mappings for two additional dangerous permissions \texttt{CAMERA} and \texttt{RECORD\_AUDIO}, which are closely related to user's privacy. 
Table~\ref{tab:permission_results} details the new permissions protected by Native-triggered permission checks excluded in the previous works. 

Due to the lack of ground truth for the permission protection mappings, it is difficult to evaluate the overall accuracy of our extracted mappings. We therefore resort to a manual process to examine the correctness of our mappings. To this end, we manually read the involved source code in the AOSP codebase to confirm if the APIs will go through the corresponding security check(s) and if the condition(s) in the security check(s) is(are) consistent with the condition(s) in the mappings. We randomly selected 10\% of the total mappings (\ie, 182 mappings) for manual inspection and found no missing or redundant protection conditions in our mappings. 


The column 2 in Table \ref{tab:performance_comparison} reports the time consumed by \tool{} to analyze the collected AOSP versions. As shown in the table, the average time taken for extracting the protection mapping from one AOSP version is 47 mins. 
Since the analysis has to be done only once per Android version, the time overhead is acceptable.


\subsection{Applications of Protection Mappings}
\label{applications_of_protection_mapping}
The protection mappings can be leveraged to detect security issues of Android apps, such as \textit{permission over-privilege} and \textit{component hijacking}. In this subsection, we evaluate the effectiveness of our extracted mappings in identifying these security vulnerabilities.

We include two categories of Android apps in our experiments: the custom ROM apps that are pre-installed on the devices, such as \textit{Camera} and \textit{Calendar}, as well as the third-party apps that users can download from official or alternative app stores (\eg, social networking apps, banking apps). The experimental dataset contains 1,035 custom ROM apps extracted from five Android custom ROMs of four vendors (\ie, Samsung, LG, Huawei, and Xiaomi) and 10,000 third-party apps randomly downloaded from the Google Play store. 
Table~\ref{tab:collect_app} shows an overview of the dataset in use. 

We use the permission protection mappings extracted from AOSP to detect security vulnerabilities in both the custom ROM apps and third-party apps. The AOSP mapping may miss some vendor-customized permissions (\eg, \texttt{huawei.permission.SET\_SMSC\_ADDRESS}), which may be used in the custom ROM apps. Nevertheless, we argue that using the AOSP mapping may miss some vulnerabilities caused by the misuse of vendor-customized permissions but will not affect the results corresponding to the official permissions, serving as the main scope of our study. 

\begin{table}[t]
\centering
\caption{Experimental dataset}
\label{tab:collect_app}
\resizebox{0.8\linewidth}{!}{
\begin{tabular}{ccrr}
\toprule
              \textbf{Set}  & \textbf{Source} & \textbf{\# Apps} \\ \midrule
\multirow{4}{*}{\textbf{Custom ROM Apps}} & \multirow{1}{*}{Xiaomi (7.0)} & 398 \\ 
& \multirow{1}{*}{LG (7.0)} & 220 \\ 
& \multirow{1}{*}{Samsung (7.0)} & 302 \\ 
& \multirow{1}{*}{Huawei (7.0)} & 115 \\ 
\textbf{Third-party Apps} & \multirow{1}{*}{Google Play Store} & 10,000\\ 
\midrule
\multirow{1}{*}{\textbf{Total}} & & 11,035 \\ 
\bottomrule 
\end{tabular}
}
\end{table}

\begin{table*}[th!]
\centering
\caption{Permission over-privilege detection results}
\label{tab:over-privilege}
\resizebox{\linewidth}{!}{
\begin{threeparttable}
\begin{tabular}{@{\extracolsep{4pt}}c@{\hspace{1ex}}c@{\hspace{1ex}}c@{\hspace{1ex}}r@{\hspace{1ex}}r@{\hspace{1ex}}r@{\hspace{1ex}}r@{\hspace{1ex}}r@{\hspace{1ex}}r@{\hspace{1ex}}r@{\hspace{1ex}}r@{}}
\toprule
\multicolumn{1}{c}{\multirow{2}{*}{\textbf{Data Set}}}  & \multicolumn{1}{c}{\multirow{2}{*}{\textbf{Source}}} &
\multicolumn{1}{c}{\multirow{2}{*}{\begin{tabular}{c}\textbf{Analyzable}\\\textbf{Apps}\end{tabular}}} & 
\multicolumn{4}{c}{\textbf{Avg \# Unneeded Permissions per App}} & \multicolumn{2}{c}{\begin{tabular}{c}\textbf{\# of False Positive}\\\textbf{Permissions}\end{tabular}} & \multicolumn{2}{c}{\begin{tabular}{c}\textbf{\% of Apps that}\\\textbf{Have False Positives}\end{tabular}} \\ 
\cline{4-7} \cline{8-9}\cline{10-11}
& & & \textbf{Nati + Ar} \tnote{1} & \textbf{Ar} & \textbf{Nati + Ax} & \textbf{Ax} & \textbf{Nati + Ar} & \textbf{Nati + Ax} & \textbf{Nati + Ar} & \textbf{Nati + Ax}\\ 
\midrule
\multirow{5}{*}{\textbf{Custom ROM Apps}} & 
\multirow{1}{*}{\textbf{Xiaomi (7.0)}} & 345 & 13.41 & 13.7 & 10.94 & 11.18 & 102 & 84 & 19.13\% & 17.97\%   \\ 
& \multirow{1}{*}{\textbf{LG (7.0)}} & 215 & 12.76 & 13.17 & 10.24 & 10.6 & 88 & 77 & 25.12\% & 24.19\%   \\ 
& \multirow{1}{*}{\textbf{Samsung (7.0)}} & 266 & 11.58 & 12.0 & 9.44 & 9.82 & 113 & 101 & 22.93\% & 21.8\%    \\ 
& \multirow{1}{*}{\textbf{Huawei (7.0)}} & 111 & 13.53 & 14.02 & 11.14 & 11.56 & 54 & 47 & 24.32\% & 22.52\%   \\ 
& \multirow{1}{*}{\textbf{Total (7.0)}} & 937 & 12.75 & 13.14 & 10.38 & 10.71 & 357 & 309 & 22.2\% & 21.02\%   \\ 
\textbf{Third-party Apps} & \multirow{1}{*}{\textbf{Google Play Store}} & 9,475 & 6.03 & 6.88 & 3.52 & 4.36 & 8,063 & 7,894 & 71.5\% & 71.38\% \\
\bottomrule 
\end{tabular}
  \begin{tablenotes}
    \item[1] \textbf{Nati}: \textsc{NatiDroid}; \textbf{Ar}: \textsc{Arcade}; \textbf{Ax}: \textsc{Axplorer}; \textbf{+}: merge the two mappings.
  \end{tablenotes}
\end{threeparttable}
}
\end{table*}

\subsubsection{Permission Over-privilege Detection}
\label{subsubsection:overprivilege}

Android app developers access Android framework functionalities by invoking Android APIs. Some APIs have access to sensitive information, such as reading the contact list, are protected by permissions. Developers need to request such permissions from the Android system before accessing the sensitive resources. Specifically, a list of required permissions need to be declared in the \texttt{AndroidManifest.xml} file, and the corresponding permissions protected APIs are to be invoked in the app's implementation. According to the Android developers' documentation~\cite{devguide}, app developers should request a minimum set of permissions required to complete the app's functionality, as introducing additional permissions will increase the risk of privacy leak. However, developers usually (either intentionally or unintentionally) request permissions that are not related to the functionalities actually implemented in the app, and hence, not necessary~\cite{felt2011android}. 

To detect apps with the permission over-privilege issue, we extract the reachable APIs of the app (with a 30-min timeout), and retrieve its protection conditions (\eg, permission, \textit{UID}, \textit{PID}) according to the mappings. For instance, when invoking an Android API, it may check the \textit{UID} (\eg, \texttt{uid == AID\_SYSTEM} checks if the app has system privilege) and the \textit{PID} (\eg, \texttt{callingPid == getpid()} examines if the method is called by its own process) along with permission enforcement.
While the \textit{UID} can be retrieved statically, the \textit{PID} has to be determined at run-time, thereby cannot be obtained through static analysis. Nevertheless, the apps included in the experiment are custom ROM apps and third-party apps, which cannot possess \textit{PID} of Android system services. Therefore, it is safe to assume that \texttt{callingPid == getpid()} will always return false in our tested apps. 
Finally, if the app declares permission (in the \texttt{AndroidManifest.xml} file) that is not required (\ie, no APIs associated with the permissions found in the app), we flag it as an over-privilege case.

\noindent\textbf{Results.~}
Table \ref{tab:over-privilege} presents the over-privilege detection results. To demonstrate the effectiveness of our mappings in pinpointing the permission over-privilege issue, we compare previous works' results  with our results. Note that, in our results (\ie, \textit{Nati + Ar} and \textit{Nati + Ax}), the Java-side mappings are derived from \textsc{Arcade} and \textsc{Axplorer}. 
We identify 95.8\% and 95.5\% apps with a permission over-privilege issue using \textsc{Arcade}'s and \textsc{Axplorer}'s mappings, respectively. Among their results, we identify that 66.6\% and 54.2\% apps (in \textsc{Arcade}'s and \textsc{Axplorer}'s results, respectively) contain false-positive results caused by missing Native-triggered permission mappings. Specifically, as shown in the last four columns of Table \ref{tab:over-privilege}, there are 8,063 and 357 permissions that are erroneously identified as over-privilege by \textsc{Axplorer} in 71.5\% third-party apps and 22.2\% custom ROM apps, respectively; for \textsc{Axplorer}, 7,894 permissions in 71.38\% third-party apps and 309 permissions in 21.02\% custom ROM apps are found to be false-positive. 

\begin{figure}[t]
  \centering                     
  \includegraphics[width=\linewidth]{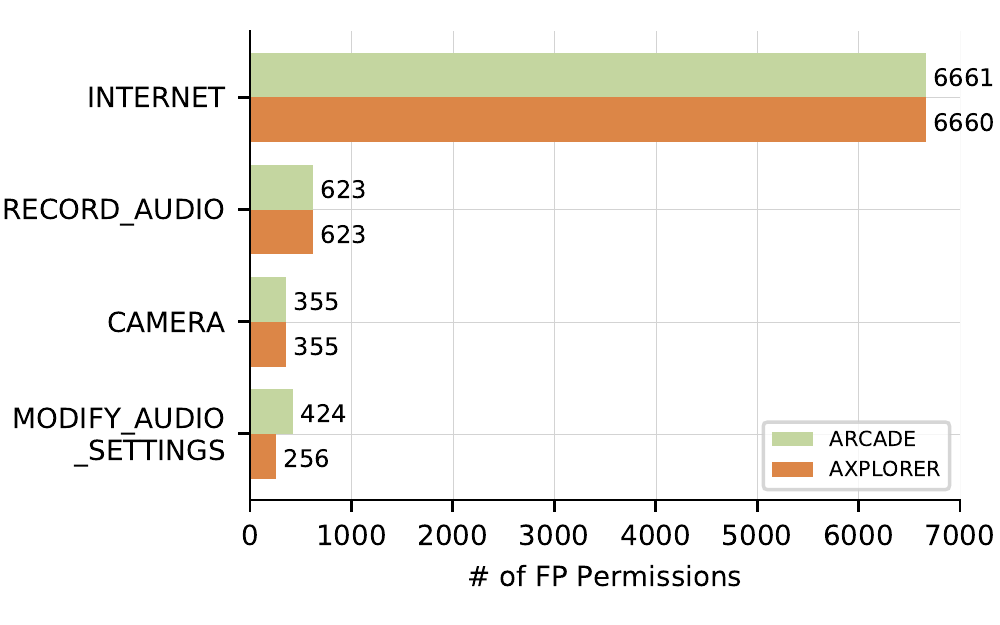}
  \caption{False-positive over-privileged permissions in previous works detected by \textsc{NatiDroid}. }
  \label{fp_permission} 
\end{figure}

Interestingly, both \textsc{Arcade} and \textsc{Axplorer} report that a significantly high proportion of apps (approximately 96\%) suffer from a permission over-privilege issue. We therefore take an in-depth look into their detection results and observe that the majority of their false positives are caused by missing native triggered \texttt{INTERNET} permission mappings. As illustrated in Figure \ref{fp_permission}, we further present the breakdown of permissions that cause the false positive results in the comparing methods. Specifically, missing \texttt{INTERNET} permission mappings leads to 6,661 and 6,660 false positives in \textsc{Arcade}'s and \textsc{Axplorer}'s results. Other missing permission mappings that contribute to the false positives include \texttt{RECORD\_AUDIO} (623 false positives in both \textsc{Arcade} and \textsc{Axplorer}), \texttt{MODIFY\_AUDIO\_SETTINGS} (424 and 256 false positives in \textsc{Arcade} and \textsc{Axplorer}, respectively), and \texttt{CAMERA} (355 false positives in both \textsc{Arcade} and \textsc{Axplorer}). 

\noindent\textbf{Manual inspection.~}Due to the lack of ground truth, we manually inspect if the over-privileged permissions detected are indeed unneeded by the containing apps. The first two authors of this paper and three security researchers are involved in the manual inspection. The result is determined via majority voting. To this end, for each app, we decompile the \textit{APK} file and locate the relevant APIs. Then, we manually check the app's context and determine whether the invocation of the APIs meet the conditions in the protection mappings. As this process involves immense manual efforts, it cannot scale to cover a large number of apps. Hence, we manually verified 100 randomly selected apps. Our manual analysis indicates that most of the cases are true positives (95\%). The remaining five apps contain implicit parameters passed to the APIs to be examined, which cannot be precisely inferred via static analysis. Nevertheless, we resort to a dynamic approach to verify the remaining five apps. 
Specifically, we remove the permissions in question from the \texttt{AndroidManifest.xml} file and repackage the app. Then, we manually test the app on an emulator to confirm if the app crashes or the corresponding functions are disabled. As a result, the removal of the permissions in question has no impact on the apps, suggesting that these permissions are indeed unneeded.

\begin{figure}[t]
    \centering
    \includegraphics[width=\linewidth]{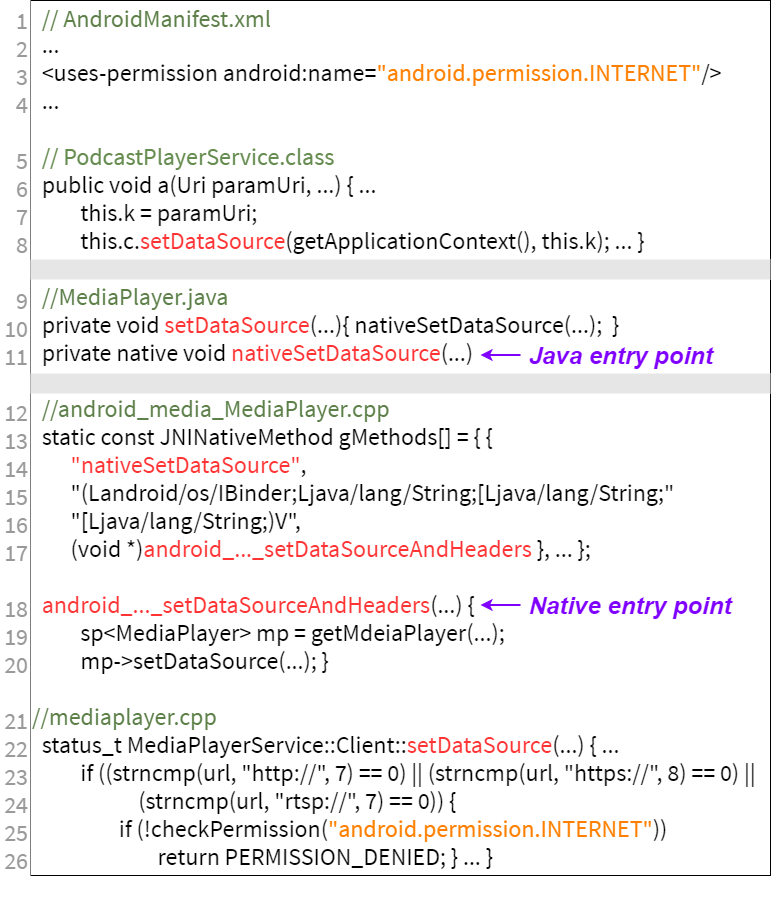}
    \caption{A case of over-privilege}
    \label{code:case_1}
\end{figure}

\noindent\textbf{Case study.~}
In our study, we find that a large number of apps misidentified as over-privilege cases by previous works because of their imprecise mapping of the \texttt{INTERNET} permission. 
We show a typical false positive case by \textsc{Axplorer} and \textsc{Arcade} in Figure \ref{code:case_1}. Lines 1 to 8 in Figure \ref{code:case_1} present the code snippet derived from the victim app \textit{Angel Numbers} \cite{angelnumbers}. The app declares the \texttt{INTERNET} permission in its manifest (line 3), and invokes the API \texttt{setDataSource()} (line 8) to play online media (hence requiring accessing the Internet). \tool{} traces the method back to its definition in the containing Java class \texttt{MediaPlayer.java} (line 10), and finds that no permission is required to access this method. Unfortunately, since the previous works \textsc{Axplorer} and \textsc{Arcade} only analyzed the Java framework but ignored the native libraries, they thereby concluded that there is no mapping between \texttt{setDataSource()} method and \texttt{INTERNET} permission. However, our approach further identify that \texttt{setDataSource()} method invokes a native method \texttt{android\_...\_setDataSourceAndHeaders()} in the native library \texttt{android\_media\_MediaPlayer.cpp} through \textit{JNI} (lines 12 to 17), which eventually triggers a permission check for \texttt{INTERNET} (line 25). By analyzing the native libraries implemented in four AOSPs, \tool{} has identified 20 mappings related to \texttt{INTERNET} permission, where the permission checks are implemented in native code.   

\subsubsection{Component Hijacking Detection}
\label{subsubsection:hijacking}

\begin{table*}[t]
\centering
\caption{Component hijacking detection results} 
\label{tab:component-hijacking}
\resizebox{\linewidth}{!}{
\begin{threeparttable}
\begin{tabular}{@{\extracolsep{4pt}}cccrrrrrrrrrr@{}}
\toprule
\multicolumn{1}{c}{\multirow{2}{*}{\textbf{Data Set}}} &  \multicolumn{1}{c}{\multirow{2}{*}{\textbf{Source}}} & 
\multicolumn{1}{c}{\multirow{2}{*}{\begin{tabular}{c}\textbf{Analyzable}\\\textbf{Apps}\end{tabular}}} & 
\multicolumn{4}{c}{\begin{tabular}{c}\textbf{\# Hijacked Components}\\\textbf{(in \# Apps)}\end{tabular}} &  \multicolumn{4}{c}{\begin{tabular}{c}\textbf{\# Hijacked Sources}\end{tabular}} & 
\multicolumn{2}{c}{\begin{tabular}{c}\textbf{\% of Apps Having}\\\textbf{ Native-triggered}\\\textbf{Component Hijacking}\end{tabular}} \\ 
\cline{4-7} \cline{8-11}\cline{12-13}
& & & \textbf{Nati + Ar}  \tnote{1} & \textbf{Ar} & \textbf{Nati + Ax} & \textbf{Ax} & \textbf{Nati + Ar} & \textbf{Ar} & \textbf{Nati + Ax} & \textbf{Ax} & \textbf{Nati + Ar} & \textbf{Nati + Ax}\\ 
\midrule
\multirow{5}{*}{\textbf{Custom ROM Apps}} & 
\multirow{1}{*}{\textbf{Xiaomi (7.0)}} & 280 & 16 (10) & 0 (0) & 16 (10) & 6 (4)  & 16  & 0  & 30  & 14  & 3.57\% & 3.57\%       \\
& \multirow{1}{*}{\textbf{LG (7.0)}} & 171 & 1 (1) & 0 (0)  & 1 (1)  & 1 (1)  & 1  & 0  & 4  & 3  & 0.58\% & 0.58\%          \\ 
& \multirow{1}{*}{\textbf{Samsung (7.0)}} & 197 & 4 (2) & 0 (0)  & 4 (2)  & 3 (1)  & 4  & 0  & 10  & 6  & 1.02\% & 1.02\%     \\ 
& \multirow{1}{*}{\textbf{Huawei (7.0)}} & 91 & 1 (1) & 0 (0)  & 1 (1)  & 1 (1)  & 2  & 0  & 7  & 5  & 1.1\% & 1.1\%         \\ 
& \multirow{1}{*}{\textbf{Total (7.0)}} & 739 & 22 (14) & 0 (0)  & 22 (14)  & 11 (7)  & 23  & 0  & 51  & 28  & 1.89\% & 1.89\%      \\ 
\textbf{Third-party Apps} & \multirow{1}{*}{\textbf{Google Play Store}} & 9,123 & 2 (2) & 0 (0)  & 4 (4)  & 3 (3)  & 2  & 0  & 8  & 6  & 0.02\% & 0.02\% \\ 
\midrule
\end{tabular}
  \begin{tablenotes}
    \item[1] \textbf{Nati}: \textsc{NatiDroid}; \textbf{Ar}: \textsc{Arcade}; \textbf{Ax}: \textsc{Axplorer}; \textbf{+}: merge the two mappings.
  \end{tablenotes}
\end{threeparttable}
}
\end{table*}

Android apps are built upon basic blocks named components, such as Activities and Services. Each component fulfills a task and can respond to requests from the app's other components and the Android framework. A component may also handle the requests from other apps if the component is publicly available (\ie, \textit{exported}). For example, a restaurant review app may need to display a map to mark the location of a restaurant. Instead of implementing its own map, the app can incorporate an existing map component exported by a navigation app.
However, suppose the component has access to sensitive information (\eg, location) but is not well-protected (\eg, through permission enforcement). In that case, it may be hijacked by other apps to gain unauthorized access to protected resources through exported components in vulnerable apps.

We identify if an app is vulnerable to component hijacking attack by comparing the set of \textit{dangerous level} permissions it requires to access an exported component (declared in \texttt{AndroidManifest.xml}) with the actual permission protected APIs accessible by the component. If the former is weaker than the latter, we consider it vulnerable to a hijacking attack. To this end, we obtain a list of control flow reachable APIs accessible by each exported component. Based on our mapping results, we can therefore infer the permissions required to protect the component (denoted as $P_{r}$). On the other hand, we retrieve the permissions enforced in the manifest for accessing these exported components (denoted as $P_{d}$). For each exported component, if $P_{r} \cap P_{d} < P_{r}$, the component is considered vulnerable to hijacking attacks.  

It should be noted that in the Android system, if a component is designed to provide a privileged functionality exclusively to other apps signed by the same developer, a signature level permission could be declared, which may not match any of the permissions in $P_{r}$. Since these components are only accessible by the apps signed with the same developer, we therefore consider the components with signature level permissions well protected, even if it satisfies the condition~$P_{r} \cap P_{d} < P_{r}$.

\noindent\textbf{Results.}
As presented in Table \ref{tab:component-hijacking}, our mapping is capable of pinpointing additional components that are vulnerable to hijacking attacks in both custom ROM and third-party apps. For instance, we have identified 24 and 12 vulnerable components in 16 apps with hijacking vulnerabilities that are overlooked by \textsc{Arcade} and \textsc{Axplorer} (columns 4 to 7 of Table~\ref{tab:component-hijacking}). Among the vulnerable components, we have discovered 25 sources (\ie, sensitive resources or operations) that may be accessed by unauthorized apps (columns 8 to 11 of Table~\ref{tab:component-hijacking}). As observed, custom ROM apps suffer more from Native-triggered component hijacking vulnerabilities than third-party apps. For instance, the apps contain Native-triggered component hijacking vulnerabilities range from 3.57\% to 0.58\% across all vendors, while there are only 2 (0.02\%) third-party apps having such a  vulnerability.

\noindent \textbf{Manual inspection.~}Similar to the manual inspection on the over-privilege detection results, two authors and three security researchers manually verified the component-hijacking results. Note that, due to the code obfuscation on 6 apps, we can only verify the 16 vulnerable components in the remaining  12 apps. For each app, we decompile the APK file and obtain the required permission in \texttt{AndroidManifest.xml} as $P_{r}$. To obtain the $P_{d}$, we start by tracing the entry-points from Java-side to their corresponding entry-points on native-side, and then verify all the Native-triggered security checks. The manual inspection finds no false positives in the component-hijacking results. However, there could be false negatives due to the incompleteness of CFGs, which is almost infeasible to be verified manually. 

\begin{figure}[t]
    \centering
    \includegraphics[width=\linewidth]{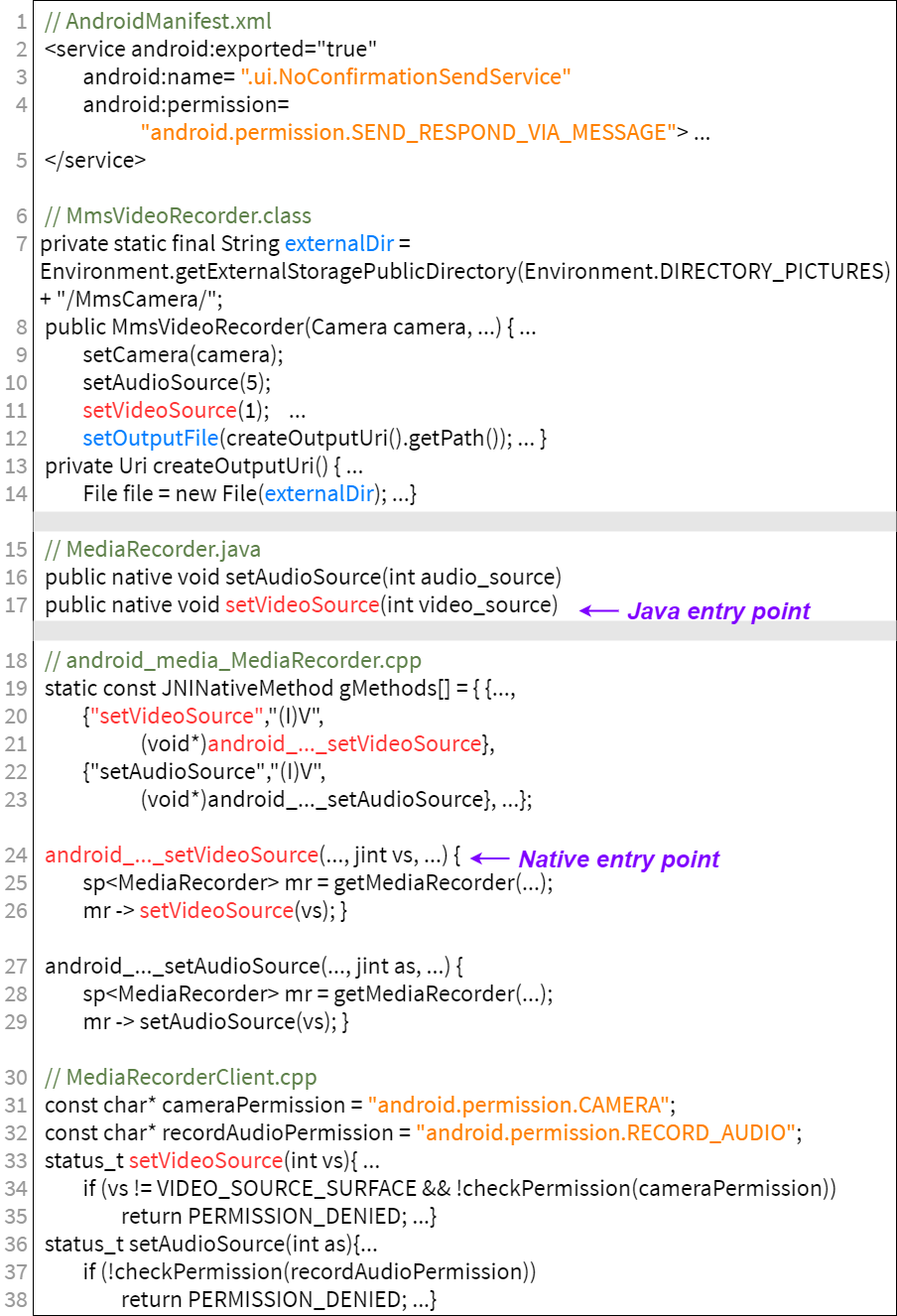}
     \vspace{1mm}
    \caption{A case of component hijacking}
    \label{code:case_2}
\end{figure}

\noindent\textbf{Case study.}
We have identified a system app \texttt{com.android.mms} in Huawei's EMUI 5.0 ROM (based on Android 7.0) with component hijacking vulnerability. Figure~\ref{code:case_2} shows the code snippet of the app (lines 1 to~14) and the involving Java framework class (lines 15 to~17) and native library (lines 18 to 38). The app \textit{exported} a service named \texttt{NoConfirmationSendService}. The service has a \textit{public} method \texttt{MmsVideoRecord()} (line 8) that records the video and audio from the device's camera and microphone. The method invokes \texttt{setVideoSource()} and \texttt{setAudioSource()} APIs, which are protected by permissions \texttt{CAMERA} (line 34) and \texttt{RECORD\_AUDIO} (line~37), respectively. These permission checks are implemented in the native library \texttt{MediaRecorderClient.cpp}, and thus cannot be captured by \textsc{Axplorer} or \textsc{Arcade}. Exporting this service offers the third-party apps capability to access the video and audio recording functions and save the recorded media to an \textit{\textbf{external directory}} which could be accessed by other apps. This leads to potential privacy leakage (lines~12 to~14). However, insufficient protection has been applied to this service, \ie, neither \texttt{CAMERA} nor \texttt{RECORD\_AUDIO} permissions are required to access the resource (as declared in line 4, only \texttt{SEND\_RESPOND\_VIA\_MESSAGE} permission is required). \texttt{SEND\_RESPOND\_VIA\_MESSAGE} is a system protected permission, which allows the app to provide instant text messages to respond to the incoming phone calls, so it is unreasonable to request \texttt{CAMERA} and \texttt{RECORD\_AUDIO} permissions. The best practice is to declare a customized permission to replace the permission group. We have disclosed 24 Native-triggered vulnerabilities to 4 vendors and 2 app developers through email on 1 June 2021, and are now in discussions with them about how severe the vulnerabilities are and the patching process. 

\section{Discussion and Limitations}

\subsection{Android Versions}
\label{Sec:discussion-android-version}
In this paper, we propose a solution to facilitate cross-language static analysis of the Android framework and build a prototype system, NatiDroid, to extract API-permission mapping from the Android OS.

To compare with the state-of-the-art works \textsc{Axplorer} and \textsc{Arcade}, which are close sourced and only generated the mappings up to Android 7.1, in our experiment, we extract the mappings from the latest versions they have (\ie, 7.0 and 7.1) and two newer versions (\ie, 8.0 and 8.1). Nevertheless, the proposed solution can apply to any Android version, with further engineering works to be done in the pre-processing module. 
Specifically, different AOSP versions uses different compiler to compile their native libraries. Hence, additional engineering efforts are required to prepare the compiling commands as described in  Section~\ref{Sec:pre-processing}.


\subsection{Custom ROMs}

Android smartphones such as Samsung, Huawei and Xiaomi, are shipped with vendor-customized Android systems (\ie, custom ROMs) rather than the AOSP. Unfortunately, these custom ROMs are not open-source. The proposed solution takes the source code as input; therefore, it cannot extract permission mappings from these close sourced custom ROMs. However, smartphone vendors can use our solution to analyze their customized Android versions based on their source code.
Nevertheless, to maintain the compatibility of running third-party apps, such custom ROMs are not likely to modify the \textit{normal} and \textit{dangerous} level permission specifications of AOSP that third-party apps can access, but rather add a few \textit{signature} level permissions for their own system apps. Therefore, the results derived from AOSP will not affect the security analysis of third-party apps, which are the majority in the Android ecosystem. On the other hand, with \tool{}, third-party vendors can perform an inner security analysis on custom ROM source code, determine whether there are errors in the implementation of permission mappings, and further detect permission over-privilege or component hijacking before releasing an app.


\subsection{Static Analysis}

When detecting over-privilege and component hijacking issues in Android apps, we may suffer from the intrinsic vulnerability of static code analysis when encountering code obfuscation, reflection, etc. These may lead to the unsoundness of our results. When building the apps' call graph for component hijacking detection, our method may yield unsound results because it may miss the context and the parameters that can only be obtained at run-time. For example, the API \texttt{android.media.MediaPlayer: void setDataSource} requires the \texttt{Internet} permission when the data source is online media. The source is not always a static string so that it may be assigned at run-time. Nevertheless, these challenges are regarded as well known and non-trivial issues to overcome in the  research community \cite{pauck2018android}. 



\section{Related Work}
\noindent\textbf{Android API protection mapping.~}
\textsc{Stowaway} \cite{felt2011android} initially explored and analyzed the Android permission specification. They extracted API mappings based on the feedback directed fuzz testing, and dynamically recorded the permission checks of APIs. The mappings they extracted are accurate, but the code coverage is limited. To address the limitations of low code coverage, \textsc{PScout} \cite{au2012pscout} uses  static analysis to extract the API protection mapping. However, they did not consider the context of the API invocation, and thus may produce false positive mappings. 
\textsc{Axplorer}~\cite{backes2016demystifying} leverages more accurate static analysis on the SDK and Android framework, and generates more precise permission specifications.  \textsc{Arcade}~\cite{aafer2018precise} conducted a similar static analysis, with additional attention  paid to extract other security mechanisms, such as \textit{UID} and \textit{PID} checks. While these works only analyzed the Java-side of the Android, none of the works has looked into the native libraries within the Android framework. Our work fills the research gap by analyzing the native libraries and their communications with the Java framework to produce more comprehensive permission protection mappings. 

\vspace{1mm}
\noindent\textbf{Cross-language analysis on Android.~}
A plethora of works have proposed to solve the analysis of cross-language code. George~\textit{et al.}~\cite{fourtounis2020identifying} scanned the binary libraries and cross-referenced the information to search the call-backs from Native code to Java. Their work focuses on the JNI mechanism alone.  However, the Android framework provides other IPC mechanisms, such as AIDL, which are not considered. 
Fengguo \textit{et al.}~\cite{wei2018amandroid} proposed a static analysis framework that focuses on performing cross-language modeling and generating call graphs for the analyzed apps. 
Nevertheless, these works are only applicable to Android apps, which are far less complicated than the Android framework we analyzed. In addition, our cross-language analysis handles various Android IPC mechanisms such as JNI and AIDL. 

\vspace{1mm}
\noindent\textbf{Static analysis on Android.~}
Static analysis is widely used in code analysis due to its fast speed and high coverage, especially in the field of Android security analysis. To date, there have been many works using static approaches to analyze the code of the Android apps. Static taint analysis can track the flow of information, detect privacy leaks and other issues, such as AndroidLeaks \cite{gibler2012androidleaks}, FlowDroid \cite{arzt2014flowdroid}, DroidSafe \cite{gordon2015information}, BidText \cite{huang2016detecting} and Amandroid \cite{wei2018amandroid}. There are also some works proposed to handle the analysis of Android inter-component communication (ICC), such as Epicc \cite{octeau2013effective}, DidFail \cite{klieber2014android} and IccTA \cite{li2014know}. Our static analysis extends existing static analysis tools to enable the cross-language analysis.


\vspace{1mm}
\noindent\textbf{Android vulnerability detection.~}
There are also many works on vulnerability identification of Android operation systems, including the leakage of content provider \cite{grace2012systematic}, data encryption vulnerabilities \cite{kim2013predictability, egele2013empirical}, cloud push-messaging services vulnerabilities \cite{li2014mayhem} and others \cite{lu2015checking,pandita2013whyper,fahl2013hey,aafer2015hare}. Specially, AutoCog~\cite{qu2014autocog} analyzed and checked that if the app's required permission conforms to the description of the permission. 
FANS~\cite{liu2020fans} proposed to fuzz the Android Native system service for detecting vulnerabilities, while 
Yousra \textit{et al.} \cite{aafer2021android} used fuzz to detect vulnerabilities in Android TVs.
By leveraging the protection mapping our approach generated, we also identified vulnerabilities in Android apps, such as permission over-privilege and component hijacking.

\section{Concluding Remarks}
We proposed a novel approach, \tool{}, to facilitate the cross-language analysis of the Android framework. \tool{} identifies the entry-point pairs of both Java- and Native-sides of the Android framework, where both sides are communicated through \textit{JNI} and \textit{AIDL} based mechanisms, so \tool{} builds the cross-language CFG on the overall Android framework (Java + Native code). Based on the cross-language CFG, we extracted Native-triggered permission specifications and created the protection mappings in the native code to complement existing Java-based mappings. We further applied our new mappings to detect \textit{permission over-privilege} and \textit{component hijacking} vulnerabilities in a large dataset consisting of more than 11,000 Android apps. 

Our results show that using the mapping derived by \tool{} can identify a significant number of false results existing in the state of the art, such as \textsc{Axplorer} and \textsc{Arcade}, notwithstanding the inevitable errors subject to the accuracy of generated CFGs. 
\ifCLASSOPTIONcaptionsoff
  \newpage
\fi



%



\bibliographystyle{unsrt}
\bibliography{permission}

%

\vspace{-10 mm} 

\begin{IEEEbiography}[{\includegraphics[width=1in,height=1.25in,clip,keepaspectratio]{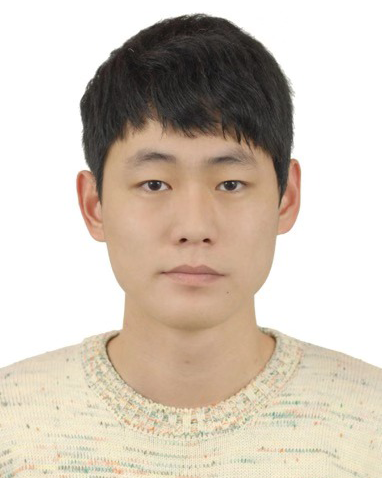}}]{Chaoran Li} received the Bachelor of Information Technology degree from Deakin University Australia in 2018. He is currently working towards the Ph.D. degree at Swinburne University of Technology. His research interests include machine learning, especially in adversarial deep learning.
\end{IEEEbiography}

\vspace{-10 mm} 

\begin{IEEEbiography}[{\includegraphics[width=1in,height=1.25in,clip,keepaspectratio]{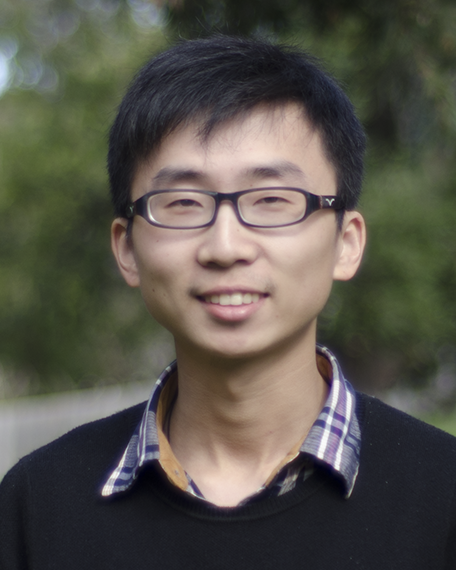}}]{Xiao Chen} is a research fellow with the Department of Software Systems and Cybersecurity, Faculty of IT, Monash University. He received Ph.D. degree from Swinburne University of Technology, Australia. His research interests include mobile software analysis, mobile security, and adversarial machine learning.
\end{IEEEbiography}

\vspace{-10 mm} 

\begin{IEEEbiography}[{\includegraphics[width=1in,height=1.25in,clip,keepaspectratio]{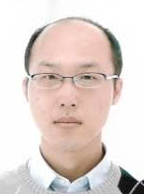}}]{Ruoxi Sun} received the Master of Computing and Innovation degree from the University of Adelaide in 2018. He is currently working towards the Ph.D. degree at the University of Adelaide. His research interests include mobile security and privacy, IoT security, and adversarial machine learning.
\end{IEEEbiography}

\vspace{-10 mm} 

\begin{IEEEbiography}[{\includegraphics[width=1in,height=1.25in,clip,keepaspectratio]{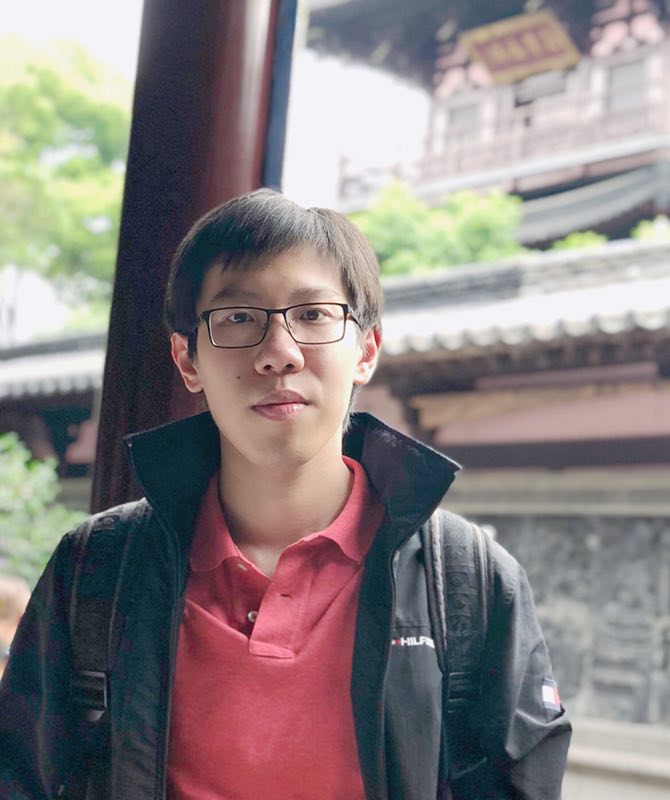}}]{Jason Xue} is a (continuing) Lecturer (a.k.a. Assistant Professor) of School of Computer Science at the University of Adelaide. He is also an Honorary Lecturer with Macquarie University. Previously, he was a Research Fellow with Macquarie University and a visiting research scientist at CSIRO-Data61 at Sydney, Australia. His current research interests are machine learning security and privacy, system and software security, and Internet measurement. He is the recipient of the ACM SIGSOFT distinguished paper award and IEEE best paper award, and his work has been featured in the mainstream press, including The New York Times, Science Daily, PR Newswire, Yahoo, The Australian Financial Review, and The Courier. He co-chaired the 1st IEEE AI4MOBILE workshop and the 1st IEEE MASS workshop on Smart City Security and Privacy. He currently serves on the Program Committee of IEEE Symposium on Security and Privacy (Oakland) 2021, ACM CCS 2021, USENIX Security 2021, 2022, NDSS 2021, 2022, IEEE/ACM ICSE 2021, 2022, PETS 2021, 2022, ESORICS 2021, and ACM ASIACCS 2021. He is a member of both ACM and IEEE.
\end{IEEEbiography}

\vspace{-10 mm} 

\begin{IEEEbiography}[{\includegraphics[width=1in,height=1.25in,clip,keepaspectratio]{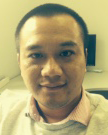}}]{Sheng Wen} received his Ph.D. degree from Deakin University, Australia, in October 2014. Currently he is a senior lecturer in Swinburne University of Technology. He has received over 3 million Australia Dollars funding from both academia and industries since 2014. He is also leading a medium-size research team in cybersecurity area. He has published more than 50 high-quality papers in the last six years in the fields of information security, epidemic modelling and source identification. His representative research outcomes have been mainly published on top journals, such as IEEE Transactions on Computers (TC), IEEE Transactions on Parallel and Distributed Systems (TPDS), IEEE Transactions on Dependable and Secure Computing (TDSC), IEEE Transactions on Information Forensics and Security (TIFS), and IEEE Communication Survey and Tutorials (CST). His research interests include social network analysis and system security.
\end{IEEEbiography}

\vspace{-10 mm} 

\begin{IEEEbiography}[{\includegraphics[width=1in,height=1.25in,clip,keepaspectratio]{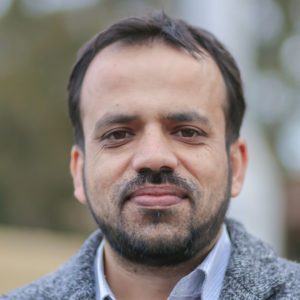}}]{Muhammad Ejaz Ahmed} received the M.S. degree in information technology from the National University of Sciences and Technology (NUST), Islamabad, Pakistan, in 2011, and the Ph.D. degree in wireless communication from the Kyung Hee University, South Korea, in 2014. From 2014 to 2015, he was a Post-Doctoral Researcher with the Pohang University of Science and Technology (POSTECH). From 2015 to 2018, he was a Research Professor with the Department of Electrical and Computer Engineering, Sungkyunkwan University (SKKU), South Korea. He is currently a Research Scientist with Data61, CSIRO, Australia. His current research interests include continuous authentication, data-driven security, malware analysis, applied machine learning, and network security.
\end{IEEEbiography}

\vspace{-10 mm} 

\begin{IEEEbiography}[{\includegraphics[width=1in,height=1.25in,clip,keepaspectratio]{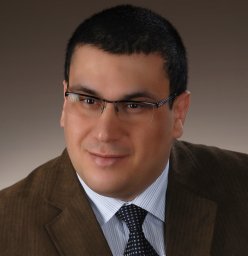}}]{Seyit Camtepe} is a senior research scientist at CSIRO Data61. He received the Ph.D. degree in computer science from Rensselaer Polytechnic Institute, New York, USA, in 2007. From 2007 to 2013, he was with the Technische Universitaet Berlin, Germany, as a Senior Researcher and Research Group Leader in Security. From 2013 to 2017, he worked as a lecturer at the Queensland University of Technology, Australia. His research interests include Pervasive security covering the topics autonomous security, malware detection and prevention, attack modelling, applied and malicious cryptography, smartphone security, IoT security, industrial control systems security, wireless physical layer security.
\end{IEEEbiography}

\vspace{-10 mm} 

\begin{IEEEbiography}[{\includegraphics[width=1in,height=1.25in,clip,keepaspectratio]{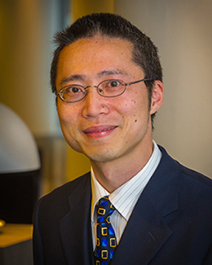}}]{Yang Xiang} received his PhD in Computer Science from Deakin University, Australia. He is currently a full professor and the Dean of Digital Research \& Innovation Capability Platform, Swinburne University of Technology, Australia. His research interests include cyber security, which covers network and system security, data analytics, distributed systems, and networking. He is also leading the Blockchain initiatives at Swinburne. In the past 20 years, he has been working in the broad area of cyber security, which covers network and system security, AI, data analytics, and networking. He has published more than 300 research papers in many international journals and conferences. He is the Editor-in-Chief of the SpringerBriefs on Cyber Security Systems and Networks. He serves as the Associate Editor of IEEE Transactions on Dependable and Secure Computing, IEEE Internet of Things Journal, and ACM Computing Surveys. He served as the Associate Editor of IEEE Transactions on Computers and IEEE Transactions on Parallel and Distributed Systems. He is the Coordinator, Asia for IEEE Computer Society Technical Committee on Distributed Processing (TCDP). He is a Fellow of the IEEE.
\end{IEEEbiography}

\end{document}